\documentclass[aps,pre,twocolumn,showpacs,superscriptaddress,10pt]{revtex4-2}

\usepackage{graphicx}
\usepackage{bm}
\usepackage{amsmath,amssymb,mathtools}
\usepackage{color}
\definecolor{darkblue}{rgb}{0,0,0.6}
\definecolor{darkred}{rgb}{0.6,0,0}
\definecolor{darkgreen}{rgb}{0,0.6,0}

\usepackage[colorlinks=true,urlcolor=darkblue,citecolor=darkblue,linkcolor=darkred,hyperfootnotes=false]{hyperref}
\makeatother

\begin{document}

\title{Thermodynamically consistent lattice Monte Carlo method for active particles}

\author{Ki-Won Kim}
\affiliation{Department of Physics and Astronomy and Center for Theoretical Physics, Seoul National University, Seoul 08826, Republic of Korea}

\author{Euijoon Kwon}
\affiliation{Department of Physics and Astronomy and Center for Theoretical Physics, Seoul National University, Seoul 08826, Republic of Korea}
\affiliation{School of Physics, Korea Institute for Advanced Study, Seoul 02455, Republic of  Korea}

\author{Yongjoo Baek}
\email{y.baek@snu.ac.kr}
\affiliation{Department of Physics and Astronomy and Center for Theoretical Physics, Seoul National University, Seoul 08826, Republic of Korea}

\begin{abstract}
Recent years have seen a growing interest in the thermodynamic cost of dissipative structures formed by active particles. Given the strong finite-size effects of such systems, it is essential to develop efficient numerical approaches that discretize both space and time while preserving the original dynamics and thermodynamics of active particles in the continuum limit. To address this challenge, we propose two thermodynamically consistent kinetic Monte Carlo methods for active lattice gases, both of which correctly reproduce the continuum dynamics. One method follows the conventional Kawasaki dynamics, while the other incorporates an extra state-dependent prefactor in the transition rate to more accurately capture the self-propulsion velocity. We find that the error scales linearly with time step size and that the state-dependent prefactor improves accuracy at high P\'{e}clet numbers by a factor of $\mathrm{Pe}^2$. Our results are supported by rigorous proof of convergence as well as extensive simulations.
\end{abstract}

\maketitle

\section{Introduction}

Active matter, a class of materials driven at the level of individual particles, has been the focus of much attention across physics, chemistry, and engineering~\cite{RamaswamyARCMP2010,MarchettiRMP2013,BechingerRMP2016,RamaswamyJSM2017,JulicherRPP2018,GompperJPCM2020,BowickPRX2022,teVrugtEPJE2025}. Among the most well-studied examples of active matter are systems of active (or self-propelled) particles, spanning a wide range of scales---from molecular motors~\cite{JulicherRMP1997}, microswimmers~\cite{ElgetiRPP2015}, and motile cells~\cite{AlertARCMP2020} to vibrated granular particles~\cite{KumarNatComms2014}, animal groups~\cite{OuellettePB2022}, and human crowds~\cite{CorbettaARCMP2023}. These systems display a multitude of fascinating nonequilibrium phenomena, such as current rectification~\cite{GalajdaJB2007,WanPRL2008,TailleurEPL2009,AngelaniPRL2009,AngelaniNJP2010,SokolovPNAS2010,DiLeonardoPNAS2010,KaiserPRL2014,MalloryPRE2014}, long-range interactions~\cite{BaekPRL2018,GranekJSM2020}, motility-induced phase separation (MIPS)~\cite{CatesARCMP2015,BialkeJNCS2015,O'ByrneBook2023}, and polar or nematic ordering~\cite{TonerAP2005,VicsekPR2012,CavagnaARCMP2014,GinelliEPJST2016,ChateARCMP2020}. Recent studies also investigated the thermodynamic aspects of such phenomena, exploring the irreversibility associated with dissipative structures formed by active particles~\cite{FodorARCMP2022}. Some of these studies are concerned with the apparent entropy production profiles, which reveal whether the observed phenomena allow for effective equilibrium descriptions~\cite{NardiniPRX2017,CrosatoPRE2019,O'ByrneNRP2022,RoPRL2022}. Other studies approach the problem on the basis of thermodynamically consistent frameworks, which explicitly model how energy flows are related to entropy production~\cite{PietzonkaJPA2018,DadhichiJSM2018,SpeckEPL2018,SpeckPRE2019,DabelowPRX2019,MarkovichPRX2021,OhPRE2023,Bebon2024}.



Active particles are intrinsically far from equilibrium, which rules out universal ensemble theories for their steady-state statistics. Additionally, nonequilibrium currents and the resultant long-range correlations lead to significantly stronger finite-size effects compared to equilibrium systems. This makes computationally efficient numerical approaches essential for studying active particle statistics. A natural solution is to develop a kinetic Monte Carlo method that discretizes time. Such methods have been used to investigate motility-induced phase separation (MIPS)~\cite{LevisPRE2014,KlamserNatComms2018}, kinetic arrest in active glasses~\cite{BerthierPRL2014}, two-dimensional melting~\cite{KlamserNatComms2018,KlamserJCP2019}, and the emergence of effective temperature~\cite{LevisEPL2015}. Although these approaches incorporated self-propulsion, they lead to an ill-defined continuous-time limit as the time step approaches zero. This limitation was addressed in \cite{KlamserPRL2021}, which introduced a class of discrete-time dynamics whose continuous-time limit recovers the prototypical active particle models, including run-and-tumble~\cite{SchnitzerPRE1993}, active Brownian~\cite{SchweitzerPRL1998}, and active Ornstein–Uhlenbeck~\cite{SzamelPRE2014,MartinPRE2021} dynamics.

To facilitate simulations across even larger length scales, computational costs can be reduced further by discretizing both space and time. Indeed, kinetic Monte Carlo methods that incorporate both discretizations have been used to study the emergence of polar order~\cite{SolonPRL2013,SolonPRE2015,BenvegnenPRL2023,WooPRL2024,ChatterjeeEPL2020,MangeatPRE2020,SolonPRL2022,ChatterjeeEPL2022,PeruaniPRL2011,RosembachPRE2024}, MIPS~\cite{ThompsonJSTAT2011,WhitelamJCP2018,PartridgePRL2019,ShiPRL2020}, and active wetting~\cite{SepulvedaPRL2017}. These studies primarily aimed to develop active lattice gas models that preserve the symmetries and conservation laws of the original continuum systems so that the universality class may remain the same. However, for investigating both thermodynamic properties and dynamical behaviors, a more robust approach is to construct a Monte Carlo method based on a thermodynamically consistent active lattice gas model. While recent proposals have introduced such models~\cite{AgranovJSTAT2022,AgranovNJP2024}, they did not address the efficiency of their numerical methods in detail.

In this study, we propose a kinetic Monte Carlo method for the active particle dynamics in discretized space and time, which ensures thermodynamic consistency and convergence to the continuum limit for both dynamic and thermodynamic quantities. We propose two possible choices for the prefactor of the transition probability, one state independent and the other state dependent. Applying the method to the steady-state behaviors of noninteracting active particles, for which the sampling error can be eliminated to reveal the error stemming solely from the discretization of space and time, we find that the error scales linearly with the time step size and that the state-dependent prefactor enhances the simulation accuracy at high P\'{e}clet number by a factor of $\mathrm{Pe}^2$. We rigorously prove these statements for various types of active particle dynamics.

The rest of the paper is organized as follows. In Sec.~\ref{sec:lattice_method}, we introduce the Monte Carlo method and the two possible choices for the prefactor. In Sec.~\ref{sec:test_method}, we apply the method to the steady-state behaviors of noninteracting active particles, demonstrating the convergence of the method to the continuum limit and quantifying how the error scales with the time step size and the P\'{e}clet number. In Sec.~\ref{sec:weak_convergence}, we prove the observed scaling behaviors of the error and justify the advantage of the state-dependent prefactor at high P\'{e}clet number. Finally, in Sec.~\ref{sec:summary}, we summarize our results and discuss possible future works.

\section{Lattice Monte Carlo Method} \label{sec:lattice_method}

Our lattice Monte Carlo method aims to simulate the dynamics of active particles in contact with a thermal reservoir at a given temperature $T$. From the perspective of a single active particle, as long as there are no alignment interactions between particles, the effects of the other particles can be represented by an external potential $V$. Thus, throughout this study, we focus on the single-particle dynamics
\begin{align}\label{eq:cont_dynamics}
	\dot{\mathbf{r}} &= \mathbf{v}-\mu\bm{\nabla} V(\mathbf{r})+\sqrt{2D}\bm{\xi},
\end{align}
where $\mathbf{r}$ is the position of the particle, $\mathbf{v}$ is its self-propulsion velocity, $\mu$ is the mobility coefficient, $D = \mu T$ is the diffusion coefficient, and $\xi$ is a white noise satisfying $\langle\bm{\xi}(t)\rangle=0$ and $\langle\xi_i(t)\xi_j(t')\rangle=\delta_{ij}\delta(t-t')$ for any $i$ and $j$. The self-propulsion $\mathbf{v}$ evolves autonomously according to its own stochastic dynamics, which do not depend on the position of the particle. For example, for the active Brownian particles (ABPs), the orientation of $\mathbf{v}$ exhibits the angular Brownian motion while keeping $|\mathbf{v}|$ constant. For the run-and-tumble particles (RTPs), the orientation of $\mathbf{v}$ is switched to a uniformly distributed random direction at a fixed rate. For the active Ornstein--Uhlenbeck particles (AOUPs), $\mathbf{v}$ follows the Ornstein--Uhlenbeck process.

Our goal is to construct a kinetic Monte Carlo method on a hypercubic lattice whose dynamics reduce to Eq.~\eqref{eq:cont_dynamics} in the continuum limit while maintaining the thermodynamic consistency between the dynamics and the energetics. This means that the probability $P(\mathbf{r}+\Delta\mathbf{r}|\mathbf{r}; \mathbf{v})$ of the active particle with the self-propulsion $\mathbf{v}$ hopping from a site $\mathbf{r}$ to a neighboring site $\mathbf{r}+\Delta\mathbf{r}$ must satisfy the local detailed balance condition~\cite{KatzJSP1984,MaesSciPost2021,PietzonkaJPA2018}
\begin{align}
	\log\frac{P(\mathbf{r}+\Delta\mathbf{r}|\mathbf{r};\mathbf{v})}{P(\mathbf{r}|\mathbf{r}+\Delta\mathbf{r};\mathbf{v})}=-\frac{\Delta Q}{T},
    \label{eq:ldb}
\end{align}
where $\Delta Q$ is the heat absorbed from the reservoir by the particle, and the Boltzmann constant is set to be unity. Note that $\mathbf{v}$ has been assumed to be even under time reversal. Generalization to the cases where $\mathbf{v}$ is odd under time reversal, although not discussed in this paper, is straightforward.

Assuming that the {\em active work} done by the self-propulsion force $\mathbf{v}/\mu$ is equal to the chemical work done by the fuel consumption (tight-coupling limit), the heat is identified by the first law of thermodynamics
\begin{align}\label{eq:heat}
	\Delta Q=\Delta V-\frac{\mathbf{v}}{\mu}\cdot\Delta\mathbf{r},
\end{align}
where $\Delta V \equiv V(\mathbf{r} + \Delta \mathbf{r}) - V(\mathbf{r})$ is the change of the particle's energy. This identification allows us to compute the reservoir entropy production associated with each hopping of the active particle, which is given by the Clausius relation
\begin{align}\label{eq:EP}
\Delta S = -\frac{\Delta Q}{T}.
\end{align}
Calculation of the system entropy production requires knowledge of the particle distribution~\cite{SeifertPRL2005}, which is a challenging task. However, if we are only interested in the steady-state or long-time behaviors of the system, the system entropy production is zero or negligible, so Eq.~\eqref{eq:EP} can be used as an estimator for the total entropy production.

To satisfy Eqs.~\eqref{eq:ldb} and \eqref{eq:heat}, the hopping probability must be of the form
\begin{align}\label{eq:trans_prob}
	P(\mathbf{r}+\Delta\mathbf{r}|\mathbf{r};\mathbf{v})
	=&C(\Delta t, \Delta\mathbf{r},\mathbf{r},\mathbf{v})
	\nonumber\\
	&\times
	\exp\left[
	\frac{1}{2D}\left(
	\mathbf{v}\cdot\Delta\mathbf{r}-\mu\Delta V
	\right)
	\right],
\end{align}
where $\Delta t$ is the length of each discretized time step, and $C$ is an undetermined prefactor satisfying $C(\Delta t, \Delta\mathbf{r},\mathbf{r},\mathbf{v})=C(\Delta t, -\Delta\mathbf{r},\mathbf{r}+\Delta \mathbf{r},\mathbf{v})$. To fix the form of $C$, let us examine under what conditions Eq.~\eqref{eq:trans_prob} achieves the correct position dynamics described by Eq.~\eqref{eq:cont_dynamics} in the continuum limit. Toward this end, we first note that the particle distribution $P(\mathbf{r},\mathbf{v},t)$ of the particle evolves according to the master equation
\begin{align} \label{eq:master_eq}
&P(\mathbf{r},\mathbf{v},t+\Delta t) = P(\mathbf{r}, \mathbf{v},t)\nonumber\\
&\qquad\qquad + \sum_{\Delta \mathbf{r}} P(\mathbf{r}|\mathbf{r}-\Delta \mathbf{r};\mathbf{v})P(\mathbf{r}-\Delta \mathbf{r},\mathbf{v},t) \nonumber\\
&\qquad\qquad - \sum_{\Delta \mathbf{r}} P(\mathbf{r}+\Delta\mathbf{r}|\mathbf{r};\mathbf{v})P(\mathbf{r},\mathbf{v},t) \nonumber\\
&\qquad\qquad + (\text{terms related to the $\mathbf{v}$ dynamics}),
\end{align}
where $\Delta \mathbf{r}$ represents all possible displacements to a neighboring site. Since the $\mathbf{v}$ dynamics is fully autonomous and unaffected by the position dynamics given by Eq.~\eqref{eq:trans_prob}, we do not need to consider them in the present discussion. Now, denoting the lattice constant by $\Delta l$, the above equation can be expanded for small $\Delta t$ and $\Delta l$ as
\begin{align} \label{eq:pos_FP}
\frac{\partial P}{\partial t} \Delta t &\simeq -\bm{\nabla}\cdot\left(\frac{\mu}{D}\tilde{\mathbf{F}}C^{(0)} + 2 \mathbf{C}^{(1)}\right)P \Delta l^2 + \nabla^2 C^{(0)} P \Delta l^2 \nonumber\\
&\quad + (\text{terms related to the $\mathbf{v}$ dynamics}),
\end{align}
where we used the simplified notations $P = P(\mathbf{r},\mathbf{v},t)$, $C^{(0)} \equiv C(0, 0,\mathbf{r},\mathbf{v})$, $\mathbf{C}^{(1)} \equiv \left.\partial C/\partial\Delta \mathbf{r}\right|_{\Delta t = 0,\,\Delta\mathbf{r}=0}$, and $\tilde{\mathbf{F}} \equiv \mathbf{v}/\mu-\bm{\nabla}V$ (note that $\partial C/\partial \Delta t$ has no contributions in the leading order). To ensure that Eqs.~\eqref{eq:cont_dynamics} and \eqref{eq:pos_FP} are consistent, the continuum limit $\Delta l \to 0$ and $\Delta t \to 0$ must be taken while keeping $\gamma^2 \equiv \Delta l^2/\Delta t$ constant, and $C$ must satisfy
\begin{align}\label{eq:cont_cond1}
	C^{(0)} = \frac{D}{\gamma^2}, \quad \mathbf{C}^{(1)} = 0.
\end{align}

This condition can be implemented in many different ways. One of the simplest examples is to choose $C$ to be equal to
\begin{align}\label{eq:C0}
	C_0\equiv \frac{\Delta t}{\Delta l^2}D,
\end{align}
which is an analog of the conventional Kawasaki dynamics~\cite{Kawasaki1972}. Alternatively, we can also choose $C$ to be equal to
\begin{align}\label{eq:Cv}
	C_v\equiv&\frac{\Delta t}{2 \Delta l^2}
	(\mathbf{v}\cdot\Delta\mathbf{r}-\mu\Delta V)
	\nonumber\\
	&\times \left\{\sinh\left[\frac{1}{2D}(\mathbf{v}\cdot\Delta\mathbf{r}-\mu\Delta V)\right]\right\}^{-1}.
\end{align}
Note that, the choice $C = C_v/(\mathbf{v}\cdot\Delta\mathbf{r}-\mu\Delta V)$ is equivalent to the conventional Glauber dynamics~\cite{GlauberJMP1963}, which was also used in the active matter context, for example, in \cite{Yao2024}. If the exerted force $\nabla V$ (satisfying $\Delta V = \nabla V \cdot \Delta\mathbf{r}$) is constant at and in the neighborhood of the site $\mathbf{r}$, the above prefactor ensures that, in addition to Eq.~\eqref{eq:cont_cond1}, the relation
\begin{align}
	\langle \Delta\mathbf{r}\rangle_{\mathbf{r},\mathbf{v}} =(\mathbf{v}-\mu\nabla V)\Delta t
\end{align}
holds exactly, where $\langle\cdot\rangle_{\mathbf{r},\mathbf{v}}$ denotes the ensemble average conditioned on the initial location $\mathbf{r}$ and the self-propulsion velocity $\mathbf{v}$. In other words, $C_v$ is chosen so that the discretized dynamics matches the actual particle velocity more closely.

With $C$ thus fixed, we propose a lattice Monte Carlo method that updates the state of the particle by the following two steps:
\begin{enumerate}
\item Determine whether the particle should hop to one of the neighboring sites according to Eq.~\eqref{eq:trans_prob}, with $C$ chosen to be $C_0$ or $C_v$.
\item Change $\mathbf{v}$ by $\Delta \mathbf{v}$ via any approximate method so that all the moments of the approximate $\Delta \mathbf{v}$ differ from the corresponding moments of the true $\Delta \mathbf{v}$ only by $\mathcal{O}(\Delta t^2)$ corrections. For example, if one knows the exact distribution of $\mathbf{v}$ after a single time step (as is the case for the ABPs, the RTPs, and the AOUPs), $\mathbf{v}$ can be updated accordingly without any error. Alternatively, if $\mathbf{v}$ follows a Langevin equation (as is the case for the ABPs and the AOUPs), one may update $\mathbf{v}$ using the Euler--Maruyama method, which ensures that the error of any moment of $\Delta\mathbf{v}$ is $\mathcal{O}(\Delta t^2)$~\cite{KloedenBook1999,AsmussenBook2007,GardinerBook2009}.
\end{enumerate}
As we demonstrate below, both $C_0$ and $C_v$ yield accurate estimations in the continuum limit. However, as it turns out, choosing $C_v$ instead of $C_0$ significantly enhances the accuracy of the simulation when the diffusion is much weaker than the drift, that is, when the P\'{e}clet number
\begin{align}\label{eq:peclet}
\mathrm{Pe} \equiv \frac{\mu|\tilde{\mathbf{F}}|\times(\text{particle size})}{D}
\end{align}
is large. Since the active nature of the particles becomes more prominent at larger Pe, we can say that $C_v$ is better than $C_0$ for simulating and exploring the unique properties of active particles.

\section{Applications of the method} \label{sec:test_method}

\begin{figure*}
\includegraphics[width=\textwidth]{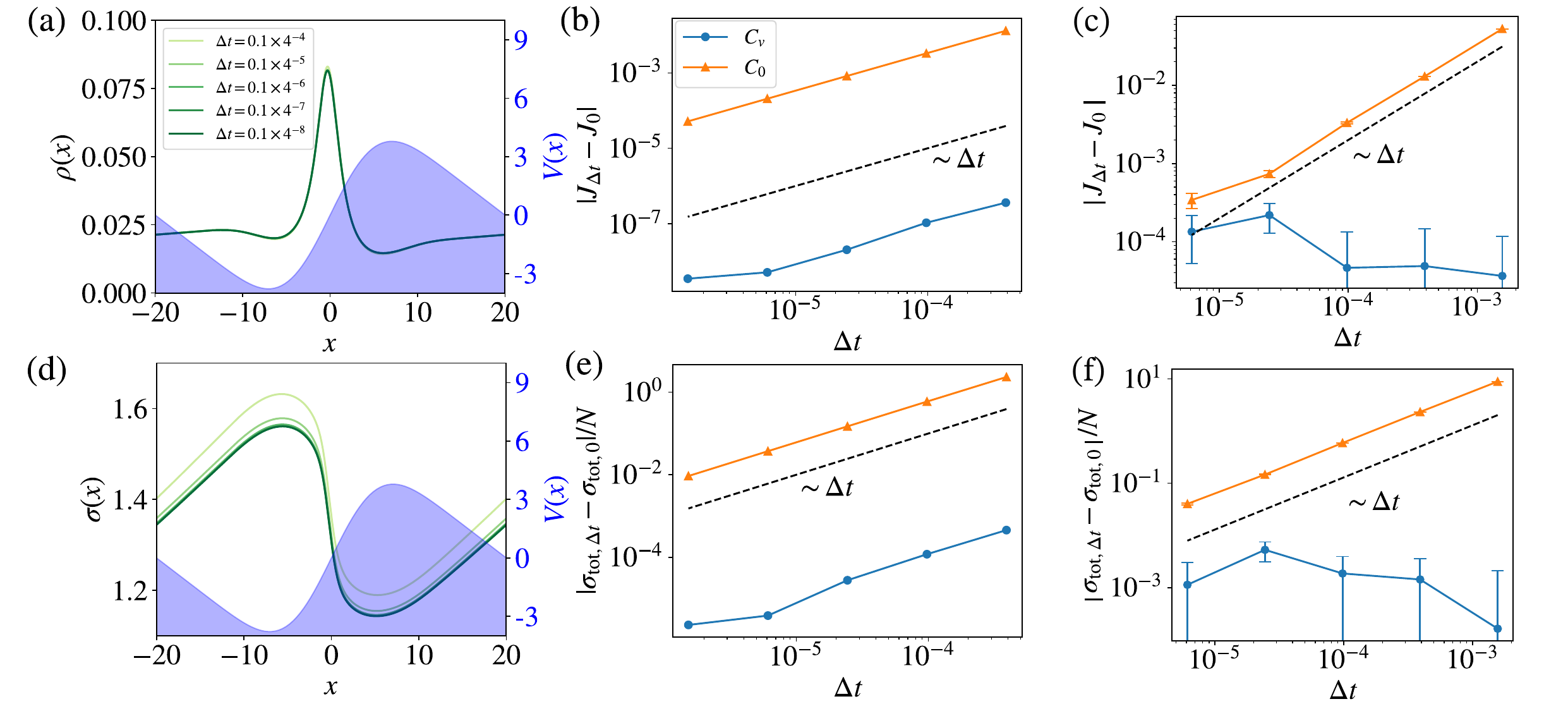}
\caption{\label{fig:current_rect} Monte Carlo simulation results of noninteracting 1D RTPs in the ratchet potential of period $L = 40$ described by Eq.~\eqref{eq:ratchet}, which is shown by shaded areas. We used $v = 1$, $\mu = 1$, $\alpha = 0.01$, $D = 0.013$, and $\Delta l^2/\Delta t = 0.625$. (a) The density profile of the RTPs (solid lines) obtained via direct integration of the probability density using $C_0$. (b) Estimated error of the steady-state current as a function of the time step size $\Delta t$, which is obtained via direct integration of the probability density using both $C_0$ and $C_v$. The continuum limit of $J_{\Delta t}$ was estimated as $J_0 \approx 0.158$. (c) Estimated error of the steady-state current with respect to the same $J_0$, which is obtained by simulating $4\times 10^5$ RTPs. (d--f) Equivalents of (a--c) for the entropy production profile $\sigma(x)$ and the total entropy production rate $\sigma_{\mathrm{tot},\Delta t}$, with the continuum limit estimated as $\sigma_{\mathrm{tot},0} \approx 53.8$.}
\end{figure*}

\subsection{Current rectification}

In the presence of an asymmetric external potential, active particles are known to achieve a nonequilibrium steady state with nonvanishing currents, a phenomenon called {\em rectification}.

As a simple system showing such behaviors, here we consider a one-dimensional (1D) system with periodic boundaries filled with noninteracting overdamped RTPs. The dynamics of each particle follows
\begin{align}
    \dot{x}(t)=vs(t)-\mu V'(x(t))+\sqrt{2D}\,\xi(t),
\end{align}
where $v$ is the self-propulsion speed, $s(t)$ the RTP polarity that switches between $+ 1$ and $-1$ at rate $\alpha$, $\mu$ the RTP mobility, $D$ the diffusion constant, and $\xi(t)$ a Gaussian white noise satisfying $\langle \xi(t) \rangle = 0$ and $\langle \xi(t)\xi(t')\rangle=\delta(t-t')$. The ratchet potential $V(x)$ is chosen as
\begin{align} \label{eq:ratchet}
    \nonumber
    V(x)=&\frac{8v}{\mu}\Bigg[\frac{7}{16}\sin\left(\frac{\pi x}{20}\right) + \frac{7}{64}\sin\left(
    \frac{\pi x}{10}\right)
    \\
    &+ \frac{1}{48}\sin\left(\frac{3\pi x}{20}\right)+\frac{1}{512}\sin\left(\frac{\pi x}{5}\right)\Bigg],
\end{align}
which has a tilted shape of period $L = 40$ as indicated by the shaded area in Fig.~\ref{fig:current_rect}(a). Note that we made the potential scale linearly with the self-propulsion force $v/\mu$ to ensure that the RTPs can easily travel across the system without being blocked by a large barrier. Moreover, since the length $L$ of the system is assumed to be dimensionless, there is no problem with making $V$ and the self-propulsion force have the same dimension.

After discretizing space and time, we simulated the system following the lattice Monte Carlo method described in Sec.~\ref{sec:lattice_method}. Since we are focusing on the case of noninteracting particles, this can be done in two different ways. First, we can simply apply the two-step procedure written above Eq.~\eqref{eq:peclet} to every particle of the system. While this method can readily be generalized to the case of interacting particles, it is also subject to sampling error since we can simulate only a finite number of particles. Since our goal here is to examine the error of the lattice Monte Carlo method stemming from the discretization of space and time (and not from the sampling error), as an alternative method, we directly simulate the probability distribution of RTPs, updating the particle distribution at each lattice point at every discretized time step following the master equation shown in Eq.~\eqref{eq:master_eq}. More specifically, the density $\rho_{+}(x,t)$ of right-movers ($s=+1$) and the density $\rho_{-}(x,t)$ of left-movers ($s=-1$) are updated according to
\begin{align}
    \nonumber
    &\rho_{\pm}(x,t+\Delta t)
    \\ \nonumber
    &=+P(x|x-\Delta l;\pm v)\rho_{\pm}(x-\Delta l,t)
    \\ \nonumber
    &\quad+P(x|x+\Delta l;\pm v)\rho_{\pm}(x+\Delta l,t)
    \\ \nonumber
    &\quad +[1-P(x+\Delta l|x;\pm v)-P(x-\Delta l|x;\pm v)]\rho_{\pm}(x,t)
    \\ 
    &\quad +\left[ -\alpha \rho_{\pm}(x,t)+\alpha \rho_{\mp}(x,t)\right]\Delta t. \label{eq:density_dyn}
\end{align}
This method amounts to simulating the system with an infinite number of particles, eliminating the sampling error via the law of large numbers. Thus, it allows us to check the error stemming from the discretization of space and time without being hindered by the sampling error.

In Fig.~\ref{fig:current_rect}(a), we show the RTP density profile in the steady state obtained by simulations using $C_0$, while varying the time step size of $\Delta t$. Clearly, the rectifying effects of the ratchet potential (shaded areas) leads to asymmetric accumulation of the RTPs (solid curves), which produces a steady-state current to the left. In this range of parameters, the finite size of $\Delta t$ has little effects on the density profile.

To examine the error more closely, we first estimate the continuum limit $J_0$ of the steady-state current by extrapolating the corresponding values $J_{\Delta t}$ measured via simulations using $C_0$ at finite $\Delta t$. This is done by fitting the simulation data to the scaling behavior $|J_{\Delta t}-J_0| \sim \Delta t$, which is later justified in Sec.~\ref{sec:weak_convergence}. Using the estimated value of $J_0$, in Fig.~\ref{fig:current_rect}(b), we compare the performances of $C_0$ and $C_v$ in terms of the error $|J_{\Delta t}-J_0|$. It turns out that the simulations using $C_v$ also exhibit $|J_{\Delta t}-J_0| \sim \Delta t$, but with a prefactor smaller by several orders of magnitude than that of $C_0$.

The density-based implementation of the Monte Carlo method via Eq.~\eqref{eq:density_dyn} is feasible only for the case where all the active particles are noninteracting, which greatly reduces the dimensionality of the problem. Thus, we also consider the particle-based implementation of the lattice Monte Carlo method, whereby we update the position of each RTP according to the probabilities given by Eq.~\eqref{eq:trans_prob}, followed by a stochastic flipping of its polarity. Here, we used the exact flipping probability corresponding to each chosen value of $\Delta t$. In this case, due to the finite number of particles, the obtained values of $J_{\Delta t}$ are subject to significant sampling fluctuations. As shown in Fig.~\ref{fig:current_rect}(c), while the data obtained using $C_0$ closely follow $|J_{\Delta t} - J_0| \sim \Delta t$, the corresponding data obtained via $C_v$ does not exhibit the same scaling behavior. This is because $|J_{\Delta t} - J_0|$ of $C_v$ has already become comparable to or much smaller than the sampling fluctuations determined by the number of particles.

The accuracies of the simulation methods can also be checked in terms of the entropy production rate. In Fig.~\ref{fig:current_rect}(d), we observe that the local entropy production rate $\sigma(x)$ is indeed nonzero everywhere across the system, reflecting the nonequilibrium nature of the steady state. Moreover, $\sigma(x)$ exhibits much stronger dependence on the size of $\Delta t$ compared to the density profile $\rho(x)$. To examine the error behaviors in more detail, we estimate the continuum limit of the total entropy production rate across the system, denoted by $\sigma_{\mathrm{tot},0}$, by extrapolating the corresponding values $\sigma_{\mathrm{tot},\Delta t }$ obtained at finite values of $\Delta t$. The method is similar to that used for estimating $J_0$. Then, in Fig.~\ref{fig:current_rect}(e), we observe that the accuracy achieved by $C_v$ is again higher than that of $C_0$ by several orders of magnitude. Since the error produced by $C_v$ is so small, if we use the particle-based implementation of the Monte Carlo method, the sampling fluctuations are the more dominant source of error than the discretization of time and space, as demonstrated by Fig.~\ref{fig:current_rect}(f).

\begin{figure}
    \includegraphics[width=\columnwidth]{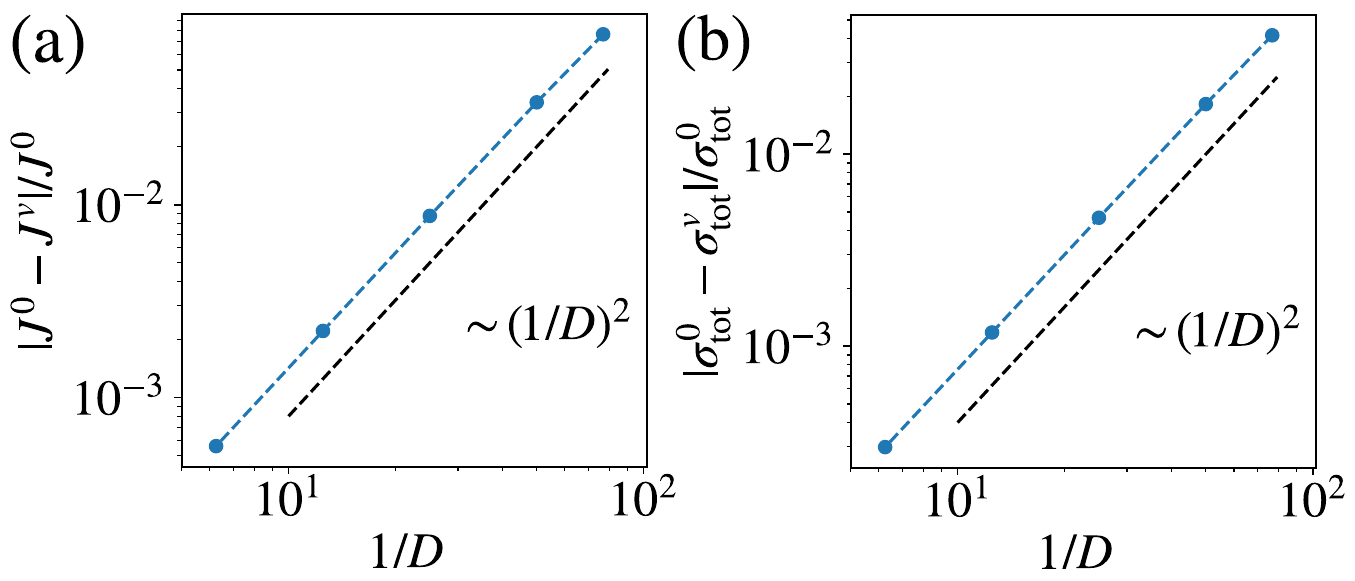}
    \caption{\label{fig:peclet} Comparison between Monte Carlo simulation results (implemented as direct integration of probability densities) of noninteracting 1D RTPs in the ratchet potential of period $L = 40$ described by Eq.~\eqref{eq:ratchet} for various values of $D$. The relative differences in the steady-state values of (a) the particle current $J$ and (b) the entropy production rate $\sigma_\mathrm{tot}$ depending on the choice of $C_0$ (which yields $J^0$ and $\sigma^0_\mathrm{tot}$) or $C_v$ (which yields $J^v$ and $\sigma^v_\mathrm{tot}$) are shown. Note that we varied the value of $D$ while keeping $\Delta t = 0.1 \times 4^{-4}$. The values of the other parameters are the same as those given in Fig.~\ref{fig:current_rect}. }
\end{figure}

All results discussed so far are obtained at a fixed value of the diffusion coefficient $D$. In Fig.~\ref{fig:peclet}, we show how the relative performances of $C_0$ and $C_v$ change as $D$ is varied, which amounts to changing the P\'{e}clet number expressed in Eq.~\eqref{eq:peclet}. We observe that $C_v$ consistently achieves higher accuracy than $C_0$ does for both the steady-state current $J$ and the entropy production rate $\sigma_\mathrm{tot}$ across the range of parameters we used, but the difference scales like $D^{-2}$. This suggests that the advantage of $C_v$ over $C_0$ is manifest in the regime of large P\'{e}clet number, {\em i.e.}, when the active particles are strongly active.

To sum up, simulations of rectified currents produced by 1D ratchet potentials demonstrate that predictions of both $C_v$ and $C_0$ converge to the same continuum limit with the error scaling as $\Delta t$, but that $C_v$ significantly outperforms $C_0$ at high P\'{e}clet number, reducing the error by a factor scaling as $\mathrm{Pe}^2$. We prove the validity of these observations in Sec.~\ref{sec:weak_convergence}, but before discussing the proof, we present another example illustrating the error behaviors.

\subsection{Pressure}

\begin{figure*}
    \includegraphics[width=\textwidth]{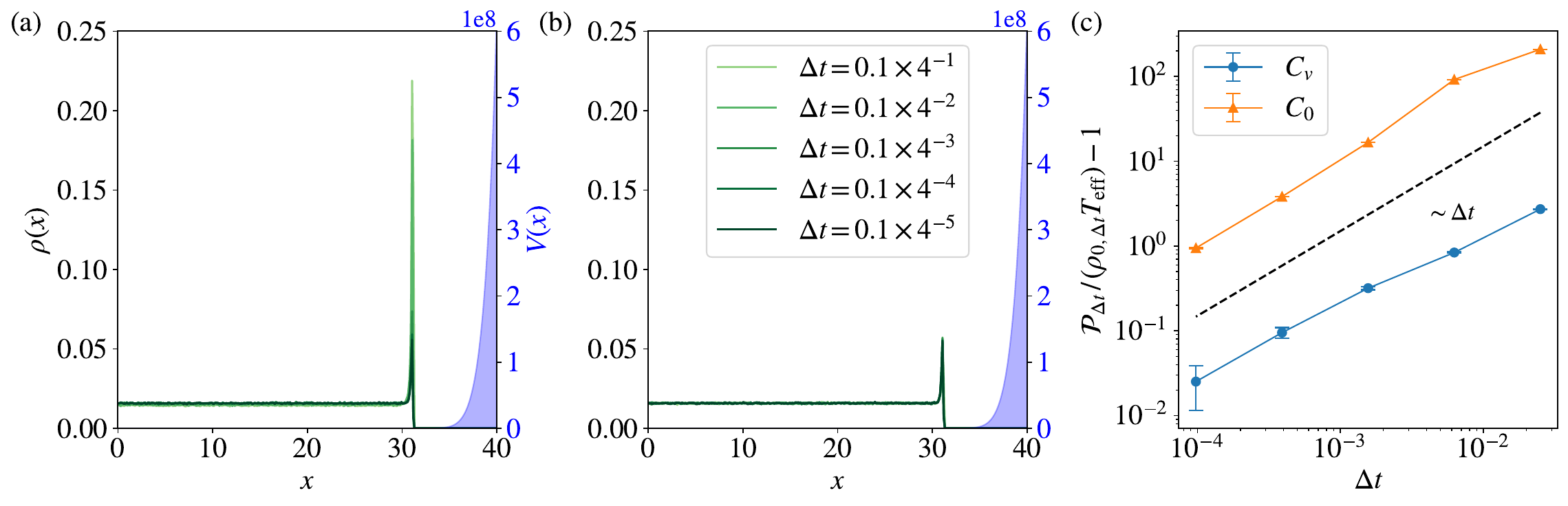}
    \caption{\label{fig:pressure} Monte Carlo simulation results of $1.5\times 10^6$ noninteracting 2D ABPs in the 2D channel with $L_y = 15$ whose walls are described by Eq.~\eqref{eq:channel}, which is shown by shaded areas. We used $v = 1$, $\mu = 1$, $D_\mathrm{r} = 1$, $D = 0.01$, and $\Delta l^2/\Delta t = 0.625$. (a) Density profiles (solid lines) of ABPs for different values of $\Delta t$ projected on the $x$-axis when $C_0$ is used. (b) The same plot obtained using $C_v$ instead of $C_0$. (c) Deviations of the estimated pressure $\mathcal{P}_{\Delta t}$ from the ideal gas law as $C_0$ or $C_v$ is used.}
\end{figure*}

In the previous example, we had to rely on extrapolation to estimate the continuum limit of physical observables since analytic solutions for the steady state were unavailable. Here we consider another example, for which the continuum limit of the observable of interest is theoretically known. The example is provided by ABPs confined in a two-dimensional (2D) channel.

Specifically, we consider a 2D channel with periodic boundaries along the $y$-direction ($0 \le y \le L_y$) and confining potentials in the $x$-direction given by
\begin{align} \label{eq:channel}
    V(x,y) = 
    \begin{cases} 
    \frac{v}{16\mu}(x-30)^8 & \text{for } x \geq 30, \\ 
    \frac{v}{16\mu}(x+30)^8 & \text{for } x \leq -30.
    \end{cases}
\end{align}
The channel is filled with noninteracting overdamped ABPs, whose dynamics are governed by
\begin{align}
    \nonumber
    &\dot{x} = v\cos\theta - \mu\,\partial_x V(x,y) + \sqrt{2D}\xi_x, \\
    \nonumber
    &\dot{y} = v\sin\theta - \mu\,\partial_y V(x,y) + \sqrt{2D}\xi_y, \\
    &\dot{\theta} = \sqrt{2D_\mathrm{r}}\eta.\label{eq:ABP}
\end{align}
Here, $v$ denotes the self-propulsion speed, $\theta$ the propulsion direction, $\mu$ the mobility, $D$ the translational diffusion coefficient, and $D_\mathrm{r}$ the rotational diffusion coefficient. Each of the independent noise components $\xi_x$, $\xi_y$, and $\eta$ is a Gaussian white noise with unit magnitude.

Let $\rho(x,y)$ denote the steady-state density profile of the ABPs. Then, the mechanical pressure applied by the ABPs on the channel wall, given by
\begin{align} 
\mathcal{P} = \frac{1}{L_y}\int^{L_y}_0 dy\,\int^\infty_0 dx\, \rho(x,y) \,\partial_x V(x,y),
\end{align}
can be calculated analytically~\cite{SolonNP2015}, yielding a variant of the ideal gas law
\begin{align} 
\mathcal{P} = \rho_0 T_\mathrm{eff} \equiv \rho_0 T\left[1 + \frac{v^2}{2D D_\mathrm{r}}\right], \label{eq:ideal_gas_law}
\end{align}
where $\rho_0$ is the bulk density. Since this relation determines the exact value of $\mathcal{P}$ in the continuum limit, it can be used to examine the error of the lattice Monte Carlo method.

In Fig.~\ref{fig:pressure}, we show the results of the particle-based implementation of the lattice Monte Carlo simulation of the 2D ABPs. More specifically, we updated the position variables $x$ and $y$ according to Eq.~\eqref{eq:trans_prob}, while the angle $\theta$ was changed by an increment $\Delta\theta$ uniformly sampled from the interval $[-\sqrt{6D_\mathrm{r}\Delta t},\,\sqrt{6D_\mathrm{r}\Delta t}]$. Simple calculations show that the error of any moment of $\Delta\theta$ scales likes $\Delta t^2$, satisfying the criterion set forth in Sec.~\ref{sec:lattice_method}. The steady-state density profiles obtained by $C_0$ and $C_v$ are shown by solid lines in Figs.~\ref{fig:pressure}(a) and (b), respectively, while the confining potentials are indicated by shaded areas. Note that the density profile and the potential are both projected on one half of the $x$-axis by averaging over the $y$-axis, since the system is homogeneous along the $y$-direction and mirror symmetric with respect to the $y$-axis. Clearly, $C_0$ exhibits much stronger finite-$\Delta t$ effects, which are manifest in the significant reduction in the number of ABPs accumulated at the walls as $\Delta t$ decreases.

This is also confirmed by how the pressure $\mathcal{P}_{\Delta t}$ measured at finite $\Delta t$ converges to the theoretical value $\mathcal{P}_0$ predicted by Eq.~\eqref{eq:ideal_gas_law}, where the bulk density $\rho_0$ is estimated as
\begin{align}
\rho_0 = \frac{1}{L_y}\int^{L_y}_0 dy\,\rho(0,y)
\end{align}
in the simulations. As shown in Fig.~\ref{fig:pressure}(c), both $C_0$ and $C_v$ yield results that converge to the correct continuum limit with the error scaling as $\Delta t$, but the prefactor of the latter is several orders of magnitude smaller than that of the former. This again demonstrates that choosing $C_v$ in place of $C_0$ at high P\'{e}clet number greatly boosts the performance of our lattice Monte Carlo method.

\section{Proof of weak convergence}
\label{sec:weak_convergence}

In this section, we prove our observations that the lattice Monte Carlo method described in Sec.~\ref{sec:lattice_method} converges to the real dynamics in the continuum limit ($\Delta t \to 0$) with the error scaling as $\Delta t$, and that using $C_v$ in place of $C_0$ reduces the error by a factor of $\mathrm{Pe}^2$ at high P\'{e}clet number. For concreteness, we consider the overdamped 2D dynamics described by
\begin{align} \label{eq:RTP}
	\dot{\mathbf{R}}(t)=v\,\mathbf{e}_\Theta(t)-\mu\,\bm\nabla V(\mathbf{R}(t))+\sqrt{2D}\,\bm{\xi}(t),
\end{align}
where $\mathbf{R}(t) \equiv (\mathbf{X}(t),\mathbf{Y}(t))$ is the position of the active particle at time $t$, $v$ is the self-propulsion strength, $\mathbf{e}_\Theta(t) \equiv (\cos\Theta(t),\sin\Theta(t))$ is the particle orientation, $\mu$ is the mobility coefficient, $V$ is the potential exerted on the particle, $D$ is the translational diffusion coefficient, and $\bm{\xi}$ is the Gaussian white noise satisfying $\langle \xi_i \rangle=0$ and $\langle \xi_i(t)\xi_j(t')\rangle=\delta_{ij}\delta(t-t')$ for any components $i$ and $j$. As for the dynamics of $\Theta$, we consider two representative models, namely the ABPs and the RTPs. For the former, $\Theta$ exhibits the Brownian motion with angular diffusion coefficient given by $D_\mathrm{r}$. For the latter, $\Theta$ switches to a uniformly distributed random value at rate $\alpha$.

To represent an arbitrary mean observable of the system, we define the function $\phi(t,\mathbf{r},\theta)$ as
\begin{align} \label{eq:phi_def}
	\phi(t,\mathbf{r},\theta) \equiv \mathbb{E}[f(\mathbf{R}(t),\Theta(t))|\mathbf{R}(0)=\mathbf{r},\,\Theta(0)=\theta],
\end{align}
where $f$ is any smooth function. By definition, we always have $\phi(0,\mathbf{r},\theta)=f(\mathbf{r},\theta)$.

Without loss of generality, we consider the case where the system evolves from $t = 0$ to $t = 1$. At the same time, we also run a lattice Monte Carlo simulation of the system, dividing the time interval into $N$ subintervals of equal length $h = 1/N$. We denote the position and the polarity of the simulated particle at the $n$th time step (at $t = t_n \equiv nh$) by $\mathbf{R}^h_n$ and $\Theta^h_n$. Meanwhile, for the sake of convenience, we also denote the position and the polarity of the actual particle governed by Eq.~\eqref{eq:RTP} by $\mathbf{R}_n \equiv \mathbf{R}(t_n)$ and $\Theta_n \equiv \Theta(t_n)$, respectively. Using these notations, we can write $\mathbf{R}_N = \mathbf{R}(1)$, $\Theta_N = \Theta(1)$, $\mathbf{R}_0 = \mathbf{R}(0) = \mathbf{r}$, and $\Theta_0 = \Theta(0) = \theta$. Moreover, since we start the actual continuum dynamics and the lattice simulation from the same initial state, we have $\mathbf{R}^h_0 = \mathbf{r}$ and $\Theta^h_0=\theta$.

For brevity, we denote by $\mathbb{E}_n[\cdot]$ the conditional average given that the state of the system at $t = t_n$ is fixed. For example, Eq.~\eqref{eq:phi_def} can also be written as
\begin{align}
    \phi(t,\mathbf{r},\theta) = \mathbb{E}_0[f(\mathbf{R}(t),\Theta(t))].
\end{align}
Then, we say that the lattice dynamics exhibit \textit{weak convergence}~\cite{KloedenBook1999,AsmussenBook2007,GardinerBook2009} to the continuum dynamics when the ``error''
\begin{align}
	\epsilon^h\equiv\mathbb{E}_0[f(\mathbf{R}^h_N,\Theta_N^h)]-\mathbb{E}_0[f(\mathbf{R}_N,\Theta_N)]
\end{align}
converges to zero as $h \to 0$. Using Eq.~\eqref{eq:phi_def}, we can also write
\begin{align}
	\epsilon^h = \mathbb{E}_0[\phi(0,\mathbf{R}^h_N,\Theta_N^h)] - \phi(1,\mathbf{r},\theta).
\end{align}

To proceed further, we define another function
\begin{align}
\psi(t,\mathbf{r},\theta) \equiv \phi(1-t,\mathbf{r},\theta),
\end{align}
which allows us to write
\begin{align}
	\epsilon^h &= \mathbb{E}_0[\psi(1,\mathbf{R}^h_N,\Theta_N^h)-\psi(0,\mathbf{R}^h_0,\Theta_0^h)] \nonumber\\
	&= \sum^{N-1}_{n=0}\mathbb{E}_0[\psi(t_{n+1},\mathbf{R}^h_{n+1},\Theta_{n+1}^h)-\psi(t_{n},\mathbf{R}^h_{n},\Theta_{n}^h)]
	\nonumber\\
	&= \sum^{N-1}_{n=0}\mathbb{E}_0[\mathbb{E}_n[\psi(t_{n+1},\mathbf{R}^h_{n+1},\Theta_{n+1}^h) - \psi(t_{n},\mathbf{R}^h_{n},\Theta_{n}^h)]]
	\nonumber\\
	&\equiv\sum^{N-1}_{n=0}\mathbb{E}_0[A^h_n],
\end{align}
where the third equality follows directly from the law of total expectation (also known as the tower rule). As proven in Appendix~\ref{sec:martingale}, $\psi(t,\mathbf{R}(t),\Theta(t))$ is a martingale, so
\begin{align} \label{eq:martingale}
    \mathbb{E}_n[\psi(t_{n+1},\mathbf{R}_{n+1},\Theta_{n+1})-\psi(t_{n},\mathbf{R}_{n},\Theta_{n})] = 0.
\end{align}
Note that the above relation holds for any choice of $\mathbf{R}_n$ and $\Theta_n$. Thus, in the following discussions, we impose that $(\mathbf{R}^h_n,\Theta^h_n)$ and $(\mathbf{R}_n,\Theta_n)$ have the same value whenever we evaluate the conditional average $\mathbb{E}_n[\cdot]$.

This allows us to reorganize $A^h_n$ as
\begin{align}
	A^h_n &=\mathbb{E}_n[\psi(t_{n+1},\mathbf{R}^h_{n+1},\Theta_{n+1}^h)-\psi(t_{n},\mathbf{R}^h_{n},\Theta_{n}^h)]
	\nonumber\\
	&\quad -\mathbb{E}_n[\psi(t_{n+1},\mathbf{R}_{n+1},\Theta_{n+1})-\psi(t_{n},\mathbf{R}_{n},\Theta_{n})] \nonumber\\
    &= B^h_n + C^h_n,
\end{align}
where
\begin{align}
	B^h_n &= \mathbb{E}_n[\psi(t_{n+1},\mathbf{R}^h_{n+1},\Theta_{n+1}^h)-\psi(t_{n+1},\mathbf{R}^h_{n+1},\Theta_{n}^h)]
	\nonumber\\
	&\quad -\mathbb{E}_n[\psi(t_{n+1},\mathbf{R}_{n+1},\Theta_{n+1})-\psi(t_{n+1},\mathbf{R}_{n+1},\Theta_{n})]
\end{align}
and
\begin{align}
	C^h_n &= \mathbb{E}_n[\psi(t_{n+1},\mathbf{R}^h_{n+1},\Theta_{n}^h)-\psi(t_{n},\mathbf{R}^h_{n},\Theta_{n}^h)]
	\nonumber\\
	&\quad -\mathbb{E}_n[\psi(t_{n+1},\mathbf{R}_{n+1},\Theta_{n})-\psi(t_{n},\mathbf{R}_{n},\Theta_{n})].
\end{align}
Among the two components of $A^h_n$, $B^h_n$ arises due to the existence of the polarity degree of freedom. Thus, for overdamped Brownian particles lacking polarity, only $C^h_n$ contributes to the error.

We start with the error contributions of $C^h_n$. Since $h = t_{n+1} - t_n$, $\Delta \mathbf{R}^h_n \equiv \mathbf{R}^h_{n+1} - \mathbf{R}^h_n$, and $\Delta \mathbf{R}_n \equiv \mathbf{R}_{n+1} - \mathbf{R}_n$ are all small, we can expand $C^h_n$ as
\begin{align} \label{eq:Chn}
	C^h_n &= \left(1+h\frac{\partial}{\partial t}+ \frac{h^2}{2}\frac{\partial^2}{\partial t^2}\right)\sum_{j=0}^\infty\sum_{k=0}^\infty \frac{1}{j!k!} \frac{\partial^{j+k} \psi}{\partial X_n^j \partial Y_n^k} E^h_{j,k}\nonumber\\
    &\quad + \mathcal{O}(h^3),
\end{align} 
where the derivatives of $\psi$ are evaluated at the state of the system at $t = t_n$ (which, as discussed above, is chosen to be the same for both lattice and continuum dynamics), and
\begin{align} \label{eq:Ehjk}
	E^h_{j,k} \equiv \mathbb{E}_n[(\Delta X^h_n)^j(\Delta Y^h_n)^k]-\mathbb{E}_n[(\Delta X_n)^j(\Delta Y_n)^k].
\end{align}
Note that $E^h_{0,0} = 0$, so the term with $j = 0$ and $k = 0$ in Eq.~\eqref{eq:Chn} does not actually contribute to $C^h_n$.

The value of $\mathbb{E}_n[(\Delta X^h_n)^j(\Delta Y^h_n)^k]$ differs depending on whether we choose $C_0$ [given in Eq.~\eqref{eq:C0}] or $C_v$ [given in Eq.~\eqref{eq:Cv}] as the prefactor $C$ of the transition probability expressed in Eq.~\eqref{eq:trans_prob}. First, let us consider the case $C = C_0$. Introducing the notations $F_X \equiv -\partial_X V$, $\tilde{F}_X \equiv v \cos(\Theta)/\mu - \partial_X V$, and $\gamma^2=\Delta l^2/\Delta t$, we can write
\begin{align}
	\label{eq:E_start}
	\mathbb{E}_n[\Delta X^h_n]
	&\simeq h \mu \tilde{F}_X
	\nonumber\\
	+h^2&\gamma^2\left(
	\frac{\mu^3 \tilde{F}_X^3}{24D^2}+\frac{\mu^2 \tilde{F}_X \partial_X F_X}{4D}+\frac{1}{6}\mu \partial_X^2 F_X
	\right),
	\nonumber \\
	\mathbb{E}_n[(\Delta X^h_n)^2]
	&\simeq 2hD+h^2\gamma^2\left(
	\frac{\mu^2 \tilde{F}_X^2}{4D}+\frac{1}{2}\mu \partial_X F_X
	\right),
\end{align}
where every function is to be evaluated at the state of the system at $t = t_n$. Since the value of $\Delta X^h_n$ can only be $\Delta l$, $0$, or $-\Delta l$, its higher-order moments are trivially related to the above two moments as
\begin{align}
\label{eq:E_start_higher}
	\mathbb{E}_n[(\Delta X^h_n)^{2k-1}]&= \Delta l^{2(k-1)}\mathbb{E}_n[\Delta X^h_n] \simeq h^k\gamma^{2(k-1)} \mu \tilde{F}_X, \nonumber \\
	\mathbb{E}_n[(\Delta X^h_n)^{2k}]&= \Delta l^{2(k-1)}\mathbb{E}_n[(\Delta X^h_n)^2] \simeq 2h^k\gamma^{2(k-1)}D
\end{align}
for $k \ge 2$. Thus, for $k>4$, we have $\mathbb{E}_n[(\Delta X^h_n)^k] = \mathcal{O}(h^3)$. The moments of $\Delta Y^h_n$ have the same forms, except that every $X$ is replaced with $Y$ and $\cos(\Theta)$ with $\sin(\Theta)$. Meanwhile, since $\Delta X^h_n$ and $\Delta Y^h_n$ cannot be simultaneously nonzero,
\begin{align}
\label{eq:E_start_covariance}
    \mathbb{E}_n[(\Delta X^h_n)^j(\Delta Y^h_n)^k] = 0
\end{align}
whenever $j$ and $k$ are both nonzero.

Now, we consider the case $C = C_v$. The lowest-order moments of $\Delta X^h_n$ are obtained as
\begin{align}
	\label{eq:E_start_v}
	\mathbb{E}_n[\Delta X^h_n] &\simeq h \mu \tilde{F}_X
	\nonumber\\
	&\quad+h^2\gamma^2\left(
	\frac{\mu^2 \tilde{F}_X \partial_X F_X}{6D}+\frac{1}{6}\mu \partial_X^2 F_X
	\right),	\nonumber\\
	\mathbb{E}_n[(\Delta X^h_n)^2]
	&\simeq 2hD+h^2\gamma^2\left(
	\frac{\mu^2 \tilde{F}_X^2}{6D}+\frac{1}{2}\mu \partial_X F_X
	\right),
\end{align}
where every function is to be evaluated at the state of the system at $t = t_n$. A comparison between Eqs.~\eqref{eq:E_start} and \eqref{eq:E_start_v} reveals some differences at order $h^2$, the most important of which concerns the order $D^{-2}$ term in $\mathbb{E}_n[\Delta X^h_n]$ which is present for $C = C_0$ but absent for $C = C_v$. The advantage of $C_v$ over $C_0$ at high P\'{e}clet number ultimately stems from this difference, as $\mathbb{E}_n[\Delta \mathbf{R}^h_n]$ is the only component of $\epsilon^h$ with $\mathcal{O}(D^{-2})$ contributions at order $h^2$, as will be confirmed by the discussions below. Meanwhile, Eqs.~\eqref{eq:E_start} and \eqref{eq:E_start_v} show no differences at order $h$. Combining this with the fact that $\Delta X^h_n$ can only be $\Delta l$, $0$, or $-\Delta l$, the leading-order behaviors of the higher-order moments are again given by Eq.~\eqref{eq:E_start_higher}. Hence, we have $\mathbb{E}_n[(\Delta X^h_n)^k] = \mathcal{O}(h^3)$ for $k > 4$. The moments of $\Delta Y^h_n$ display the same behaviors, albeit with every $X$ replaced with $Y$ and $\cos(\Theta)$ with $\sin(\Theta)$. Also, Eq.~\eqref{eq:E_start_covariance} still holds in this case.

Meanwhile, for the continuum dynamics, the evolution of $X_n$ between $t_n$ and $t_{n+1}$ can be expressed as
\begin{align}\label{eq:evol_X}
	\Delta X_n
	&=v\int^{t_{n+1}}_{t_n} d\tau\,\cos\Theta(\tau) +\mu F_X h+\sqrt{2D}\Delta W_{X,n}
	\nonumber\\
	&\quad +\sqrt{2D}\mu\bm{\nabla}F_X\cdot\int^{t_{n+1}}_{t_n} d\tau\int^\tau_{t_n} d\mathbf{W}(\tau')
	\nonumber\\
	&\quad +\frac{1}{2}(\mu^2 \bm{\nabla}F_X\cdot \tilde{\mathbf{F}}+D\mu \nabla^2 F_X)h^2+\mathcal{O}(h^{\frac{5}{2}}),
\end{align}
where $W_X(t)$ and $W_Y(t)$ are two independent Wiener processes with $\mathbf{W}(t) = (W_X(t),W_Y(t))$, $\Delta W_{X,n} \equiv W_X(t_{n+1})-W_X(t_n)$, and $\Delta W_{Y,n} \equiv W_Y(t_{n+1})-W_Y(t_n)$. The corresponding expression for $\Delta Y_n$ is a close analog of Eq.~\eqref{eq:evol_X}, except for $\cos\Theta(\tau)$ replaced with $\sin\Theta(\tau)$ and the exchange of $X$ and $Y$. From these results, we obtain the first moments
\begin{align}
	\mathbb{E}_n[\Delta X_n]
	&\simeq h\mu \tilde{F}_X\nonumber\\
    +\frac{1}{2}&h^2\left(-\frac{v}{\tau_\mathrm{p}} \cos \Theta_N +\mu^2 \bm{\nabla}F_X \cdot \tilde{\mathbf{F}}+D\mu \nabla^2 F_X\right), \nonumber\\ 
    \mathbb{E}_n[\Delta Y_n]
	&\simeq h\mu \tilde{F}_Y\nonumber\\
    +\frac{1}{2}&h^2\left(-\frac{v}{\tau_\mathrm{p}}  \sin \Theta_N +\mu^2 \bm{\nabla}F_Y \cdot \tilde{\mathbf{F}}+D\mu \nabla^2 F_Y\right),
\end{align}
where $\tau_\mathrm{p}$ is the persistence time given by $\alpha^{-1}$ for the RTPs and $D_\mathrm{r}^{-1}$ for the ABPs. Proceeding further, the second moments are obtained as
\begin{align}
	\mathbb{E}_n[(\Delta X_n)^2]
	&\simeq 2hD+h^2\mu^2 \tilde{F}_X^2, \nonumber\\
    \mathbb{E}_n[(\Delta Y_n)^2]
	&\simeq 2hD+h^2\mu^2 \tilde{F}_Y^2, \nonumber\\
    \mathbb{E}_n[\Delta X_n \Delta Y_n] &\simeq
    h^2[\tilde{F}_X\tilde{F}_Y+\mu D(\partial_X F_Y+\partial_Y F_X )],
\end{align}
 Since $\Delta X_n$ and $\Delta Y_n$ are Gaussian random variables at the leading order, all higher-order moments can be expressed as the products of the moments obtained above. Thus, $\mathbb{E}_n[(\Delta X_n)^j(\Delta Y_n)^k] = \mathcal{O}(h^2)$ whenever $j+k \ge 3$. Also note that none of these moments contain $\mathcal{O}(D^{-1})$ terms at order $h^2$. In other words, the $\mathcal{O}(D^{-1})$ components of the error are entirely artifacts of the lattice Monte Carlo method.

Collecting all the results and using them in Eq.~\eqref{eq:Ehjk}, it is clear that $E^h_{j,k}=\mathcal{O}(h^2)$ for every $j$ and $k$ regardless of whether we choose $C = C_0$ or $C = C_v$ in the transition probability. Then, since the error $\epsilon^h$ consists of the sum over $N = 1/h$ instances of such $C^h_n$'s, the total contribution of $C^h_n$ to the error is of order $h$.

Next, we move on to the error contributions of $B^h_n$. Since the angular dynamics do not depend on the position dynamics, we have
\begin{align}
    \mathbb{E}_n[g_1(\Theta_{n+1}^h)g_2(\mathbf{R}^h_{n+1})] &= \mathbb{E}_n[g_1(\Theta_{n+1}^h)]\,\mathbb{E}_n[g_2(\mathbf{R}^h_{n+1})], \nonumber\\
    \mathbb{E}_n[g_1(\Theta_{n+1})g_2(\mathbf{R}_{n+1})] &= \mathbb{E}_n[g_1(\Theta_{n+1})]\,\mathbb{E}_n[g_2(\mathbf{R}_{n+1})]
\end{align}
for arbitrary functions $g_1$ and $g_2$. Thus, $B^h_n$ can be expanded as
\begin{align}
    B^h_n
	&= \left(1+h\frac{\partial}{\partial t}+\frac{h^2}{2}\frac{\partial^2}{\partial t^2}\right)\nonumber\\
    &\quad  \sum_{j=0}^\infty\sum_{k=0}^\infty 
    \Bigg\{
    \mathbb{E}_n\!\Bigg[
    \left.\frac{\partial^{j+k} \psi}{\partial X_n^j \partial Y_n^k}\right|^{\Theta_{n+1}^h}_{\Theta_{n}^h}\Bigg] \mathbb{E}_n[(\Delta X^h_n)^j(\Delta Y^h_n)^k]
    \nonumber\\
    &\quad -\mathbb{E}_n\!\Bigg[
    \left.\frac{\partial^{j+k} \psi}{\partial X_n^j \partial Y_n^k}\right|^{\Theta_{n+1}}_{\Theta_{n}}\Bigg] \mathbb{E}_n[(\Delta X_n)^j(\Delta Y_n)^k]
    \Bigg\}
    \nonumber\\
    &\quad+ \mathcal{O}(h^3),
\end{align}
where we used the notation $\left. g\right|^{\theta_2}_{\theta_1} \equiv g(t_n,\mathbf{R}_n,\theta_2)-g(t_n,\mathbf{R}_n,\theta_1)$ for an arbitrary function $g$ and the angles $\theta_1$ and $\theta_2$.

In the case $j = k = 0$, the summand is given by
\begin{align}
    &\mathbb{E}_n[\psi(\Theta^h_{n+1})-\psi(\Theta_{n+1})]\nonumber\\ &\quad =\sum^\infty_{j=1}\frac{1}{j!}\frac{\partial^{j} \psi}{\partial\Theta^j}\,\mathbb{E}_n[(\Delta\Theta^h_n)^j-(\Delta\Theta_n)^j],
\end{align}
with $\Delta\Theta^h_n\equiv\Theta^h_{n+1}-\Theta^h_{n}$ and $\Delta\Theta_n\equiv\Theta_{n+1}-\Theta_{n}$. If the angle dynamics satisfy the error criterion set forth in Sec.~\ref{sec:lattice_method}, the above summand is of order $h^2$.

Meanwhile, in the case $j + k > 0$, according to the previous discussions, $\mathbb{E}_n[(\Delta X^h_n)^j(\Delta Y^h_n)^k] = \mathcal{O}(h)$ and $\mathbb{E}_n[(\Delta X_n)^j(\Delta Y_n)^k] = \mathcal{O}(h)$. Moreover, as long as the change of the angle distribution over a single time step is order $h$, both $\mathbb{E}_n\!\left[
    \left.\partial^{j+k} \psi/(\partial X_n^j \partial Y_n^k)\right|^{\Theta_{n+1}^h}_{\Theta_{n}^h}\right]$ and $\mathbb{E}_n\!\left[
    \left.\partial^{j+k} \psi/(\partial X_n^j \partial Y_n^k)\right|^{\Theta_{n+1}}_{\Theta_{n}}\right]$ are order $h$.

Thus, we have 
\begin{align}
	B^h_n = \mathcal{O}(h^2),
\end{align}
whose total contribution to $\epsilon^h$ is of order $h$ with no difference between $C_0$ and $C_v$ in the leading order.

This proves for the 2D RTPs and ABPs that the lattice Monte Carlo method weakly converges to the continuum dynamics, with the error scaling linearly with the time step $\Delta t = h$. Moreover, the proof also shows that choosing $C_v$ instead of $C_0$ removes the order $D^{-2}$ component of the error, enhancing the accuracy of the lattice Monte Carlo method at high P\'{e}clet number.

Our proof can readily be extended to 1D and three or higher-dimensional systems. Moreover, while we have focused on the RTPs and the ABPs for concreteness, the same proof also works for any unbiased Markovian dynamics of $\Theta(t)$. In fact, $\Theta(t)$ does not even have to be an angle---a similar proof, in which $v \cos\Theta(t)$ and $v \sin\Theta(t)$ are replaced with appropriate random variables, would also apply to the AOUPs. Hence, we expect the weak convergence of our lattice Monte Carlo method and the accuracy enhancement by $C_v$ to be valid for a broad range of active particles whose angular dynamics are unbiased and independent of position.

\section{Summary and outlook} \label{sec:summary}

In this work, we proposed a kinetic Monte Carlo method that incorporates thermodynamic consistency as well as space and time discretization, offering an efficient numerical approach for studying both dynamics and energetics of active particle systems. Our method applies to any overdamped active particle whose position follows the Langevin dynamics and the self-propulsion force evolves autonomously without depending on the position degrees of freedom. We proved that our method yields the correct continuum limit for any observable that can be expressed as the mean value of a function of the particle's state, with the error scaling linearly with the time step size $\Delta t$ as we approach the continuum limit. Furthermore, we proved that the error can be reduced by a factor of $\mathrm{Pe}^2$ at high P\'{e}clet number if we introduce a suitable state dependence in the prefactor of the transition probability instead of assuming the conventional Kawasaki dynamics. These results were explicitly demonstrated using the rectified current and the pressure of noninteracting active particles.

The key significance of our method lies in its ability to yield accurate estimations of thermodynamic quantities in the continuum limit, especially when the system is strongly active (by having a high P\'{e}clet number). This can open up new avenues for exploring the energy dissipation profiles of engines and dissipative structures formed by active particles. While this study focused on the steady-state behaviors of noninteracting active particles to demonstrate the convergence to the continuum limit, future studies will address the applications to the interacting active particles, investigating the thermodynamic aspects of both steady-state and transient collective phenomena.

{\em Code availability.}---The computer codes used to generate the results of this paper are available at \cite{code}.

\begin{acknowledgments}
This work was supported by the National Research Foundation of Korea (NRF) grants funded by the Korea government (MSIT) (RS-2023-00278985, RS-2024-00410147) and by Creative-Pioneering Researchers Program through Seoul National University.
\end{acknowledgments}
 
\appendix

\section{Proof of martingale property}
\label{sec:martingale}

Here we present a proof of the martingale property expressed in Eq.~\eqref{eq:martingale}, which can be shown for any homogeneous Markov process $\mathbf{X}_t$ with $0 \le t \le 1$. Generalizing the definition of $\psi$ to this case, we can write
\begin{align} \label{eq:psi_gen}
    \psi(t,\mathbf{x}) &\equiv \int d\mathbf{X}_t\, f(\mathbf{X}_t)P[\mathbf{X}_t,1-t|\mathbf{x},0] \nonumber\\
    &=\int d\mathbf{X}_t \,f(\mathbf{X}_t)P[\mathbf{X}_t,1|\mathbf{x},t],
\end{align}
where the second equality holds due to the homogeneity of the process.

For $t<t'\le 1$, we have
\begin{align}
    &\mathbb{E}[\psi(t',\mathbf{X}_{t'})|\mathbf{X}_t]
    \nonumber\\
    &=\int d\mathbf{X}_{t'}\,\psi(t',\mathbf{X}_{t'})P{[\mathbf{X}_{t'},t'|\mathbf{X}_t,t]}\nonumber\\
    &=\int d\mathbf{X}_{t'}\int d\mathbf{x}'f(\mathbf{x}')P[\mathbf{x}',1|\mathbf{X}_{t'},t']P{[\mathbf{X}_{t'},t'|\mathbf{X}_t,t]},
\end{align}
where the second equality comes from Eq.~\eqref{eq:psi_gen}. 

Since $\mathbf{X}_t$ is Markovian, we can use the Chapman--Kolmogorov equation
\begin{align}
    \int d\mathbf{X}_{t'}\,P[\mathbf{x}',1|\mathbf{X}_{t'},t']P{[\mathbf{X}_{t'},t'|\mathbf{X}_t,t]} = P[\mathbf{x}',1|\mathbf{X}_t,t],
\end{align}
which implies
\begin{align}
    \mathbb{E}[\psi(t',\mathbf{X}_{t'})|\mathbf{X}_t]
    &=\int d\mathbf{x}' f(\mathbf{x}')P[\mathbf{x}',1|\mathbf{X}_t,t]\nonumber\\
    &=\psi(t,\mathbf{X}_t) = \mathbb{E}[\psi(t,\mathbf{X}_t)|\mathbf{X}_t].
\end{align}
Thus, we have
\begin{align}
    \mathbb{E}[\psi(t',\mathbf{X}_{t'})-\psi(t,\mathbf{X}_t)|\mathbf{X}_t]=0,
\end{align}
confirming that $\psi(t,\mathbf{X}_t)$ is a martingale. Note that Eq.~\eqref{eq:martingale} is just a special case of the above relation.

\bibliography{lattice.bib}

\begin{thebibliography}{84}%
\makeatletter
\providecommand \@ifxundefined [1]{%
 \@ifx{#1\undefined}
}%
\providecommand \@ifnum [1]{%
 \ifnum #1\expandafter \@firstoftwo
 \else \expandafter \@secondoftwo
 \fi
}%
\providecommand \@ifx [1]{%
 \ifx #1\expandafter \@firstoftwo
 \else \expandafter \@secondoftwo
 \fi
}%
\providecommand \natexlab [1]{#1}%
\providecommand \enquote  [1]{``#1''}%
\providecommand \bibnamefont  [1]{#1}%
\providecommand \bibfnamefont [1]{#1}%
\providecommand \citenamefont [1]{#1}%
\providecommand \href@noop [0]{\@secondoftwo}%
\providecommand \href [0]{\begingroup \@sanitize@url \@href}%
\providecommand \@href[1]{\@@startlink{#1}\@@href}%
\providecommand \@@href[1]{\endgroup#1\@@endlink}%
\providecommand \@sanitize@url [0]{\catcode `\\12\catcode `\$12\catcode `\&12\catcode `\#12\catcode `\^12\catcode `\_12\catcode `\%12\relax}%
\providecommand \@@startlink[1]{}%
\providecommand \@@endlink[0]{}%
\providecommand \url  [0]{\begingroup\@sanitize@url \@url }%
\providecommand \@url [1]{\endgroup\@href {#1}{\urlprefix }}%
\providecommand \urlprefix  [0]{URL }%
\providecommand \Eprint [0]{\href }%
\providecommand \doibase [0]{https://doi.org/}%
\providecommand \selectlanguage [0]{\@gobble}%
\providecommand \bibinfo  [0]{\@secondoftwo}%
\providecommand \bibfield  [0]{\@secondoftwo}%
\providecommand \translation [1]{[#1]}%
\providecommand \BibitemOpen [0]{}%
\providecommand \bibitemStop [0]{}%
\providecommand \bibitemNoStop [0]{.\EOS\space}%
\providecommand \EOS [0]{\spacefactor3000\relax}%
\providecommand \BibitemShut  [1]{\csname bibitem#1\endcsname}%
\let\auto@bib@innerbib\@empty
\bibitem [{\citenamefont {Ramaswamy}(2010)}]{RamaswamyARCMP2010}%
  \BibitemOpen
  \bibfield  {author} {\bibinfo {author} {\bibfnamefont {S.}~\bibnamefont {Ramaswamy}},\ }\bibfield  {title} {\bibinfo {title} {{The Mechanics and Statistics of Active Matter}},\ }\href {https://doi.org/10.1146/annurev-conmatphys-070909-104101} {\bibfield  {journal} {\bibinfo  {journal} {Annu. Rev. Condens. Matter Phys.}\ }\textbf {\bibinfo {volume} {1}},\ \bibinfo {pages} {323 } (\bibinfo {year} {2010})}\BibitemShut {NoStop}%
\bibitem [{\citenamefont {Marchetti}\ \emph {et~al.}(2013)\citenamefont {Marchetti}, \citenamefont {Joanny}, \citenamefont {Ramaswamy}, \citenamefont {Liverpool}, \citenamefont {Prost}, \citenamefont {Rao},\ and\ \citenamefont {Simha}}]{MarchettiRMP2013}%
  \BibitemOpen
  \bibfield  {author} {\bibinfo {author} {\bibfnamefont {M.~C.}\ \bibnamefont {Marchetti}}, \bibinfo {author} {\bibfnamefont {J.~F.}\ \bibnamefont {Joanny}}, \bibinfo {author} {\bibfnamefont {S.}~\bibnamefont {Ramaswamy}}, \bibinfo {author} {\bibfnamefont {T.~B.}\ \bibnamefont {Liverpool}}, \bibinfo {author} {\bibfnamefont {J.}~\bibnamefont {Prost}}, \bibinfo {author} {\bibfnamefont {M.}~\bibnamefont {Rao}},\ and\ \bibinfo {author} {\bibfnamefont {R.~A.}\ \bibnamefont {Simha}},\ }\bibfield  {title} {\bibinfo {title} {{Hydrodynamics of soft active matter}},\ }\href {https://doi.org/10.1103/revmodphys.85.1143} {\bibfield  {journal} {\bibinfo  {journal} {Rev. Mod. Phys.}\ }\textbf {\bibinfo {volume} {85}},\ \bibinfo {pages} {1143} (\bibinfo {year} {2013})}\BibitemShut {NoStop}%
\bibitem [{\citenamefont {Bechinger}\ \emph {et~al.}(2016)\citenamefont {Bechinger}, \citenamefont {Di~Leonardo}, \citenamefont {L\"{o}wen}, \citenamefont {Reichhardt}, \citenamefont {Volpe},\ and\ \citenamefont {Volpe}}]{BechingerRMP2016}%
  \BibitemOpen
  \bibfield  {author} {\bibinfo {author} {\bibfnamefont {C.}~\bibnamefont {Bechinger}}, \bibinfo {author} {\bibfnamefont {R.}~\bibnamefont {Di~Leonardo}}, \bibinfo {author} {\bibfnamefont {H.}~\bibnamefont {L\"{o}wen}}, \bibinfo {author} {\bibfnamefont {C.}~\bibnamefont {Reichhardt}}, \bibinfo {author} {\bibfnamefont {G.}~\bibnamefont {Volpe}},\ and\ \bibinfo {author} {\bibfnamefont {G.}~\bibnamefont {Volpe}},\ }\bibfield  {title} {\bibinfo {title} {{Active particles in complex and crowded environments}},\ }\href {https://doi.org/10.1103/RevModPhys.88.045006} {\bibfield  {journal} {\bibinfo  {journal} {Rev. Mod. Phys.}\ }\textbf {\bibinfo {volume} {88}},\ \bibinfo {pages} {045006} (\bibinfo {year} {2016})}\BibitemShut {NoStop}%
\bibitem [{\citenamefont {Ramaswamy}(2017)}]{RamaswamyJSM2017}%
  \BibitemOpen
  \bibfield  {author} {\bibinfo {author} {\bibfnamefont {S.}~\bibnamefont {Ramaswamy}},\ }\bibfield  {title} {\bibinfo {title} {{Active matter}},\ }\href {https://doi.org/10.1088/1742-5468/aa6bc5} {\bibfield  {journal} {\bibinfo  {journal} {J. Stat. Mech.: Theor. Exp.}\ }\textbf {\bibinfo {volume} {2017}},\ \bibinfo {pages} {054002} (\bibinfo {year} {2017})}\BibitemShut {NoStop}%
\bibitem [{\citenamefont {J\"{u}licher}\ \emph {et~al.}(2018)\citenamefont {J\"{u}licher}, \citenamefont {Grill},\ and\ \citenamefont {Salbreux}}]{JulicherRPP2018}%
  \BibitemOpen
  \bibfield  {author} {\bibinfo {author} {\bibfnamefont {F.}~\bibnamefont {J\"{u}licher}}, \bibinfo {author} {\bibfnamefont {S.~W.}\ \bibnamefont {Grill}},\ and\ \bibinfo {author} {\bibfnamefont {G.}~\bibnamefont {Salbreux}},\ }\bibfield  {title} {\bibinfo {title} {{Hydrodynamic theory of active matter}},\ }\href {https://doi.org/10.1088/1361-6633/aab6bb} {\bibfield  {journal} {\bibinfo  {journal} {Rep. Prog. Phys.}\ }\textbf {\bibinfo {volume} {81}},\ \bibinfo {pages} {076601} (\bibinfo {year} {2018})}\BibitemShut {NoStop}%
\bibitem [{\citenamefont {Gompper}\ \emph {et~al.}(2020)\citenamefont {Gompper}, \citenamefont {Winkler},\ and\ \citenamefont {{Speck {\em et al.}}}}]{GompperJPCM2020}%
  \BibitemOpen
  \bibfield  {author} {\bibinfo {author} {\bibfnamefont {G.}~\bibnamefont {Gompper}}, \bibinfo {author} {\bibfnamefont {R.~G.}\ \bibnamefont {Winkler}},\ and\ \bibinfo {author} {\bibfnamefont {T.}~\bibnamefont {{Speck {\em et al.}}}},\ }\bibfield  {title} {\bibinfo {title} {{The 2020 motile active matter roadmap}},\ }\href {https://doi.org/10.1088/1361-648x/ab6348} {\bibfield  {journal} {\bibinfo  {journal} {J. Phys.: Conden. Matter}\ }\textbf {\bibinfo {volume} {32}},\ \bibinfo {pages} {193001} (\bibinfo {year} {2020})}\BibitemShut {NoStop}%
\bibitem [{\citenamefont {Bowick}\ \emph {et~al.}(2022)\citenamefont {Bowick}, \citenamefont {Fakhri}, \citenamefont {Marchetti},\ and\ \citenamefont {Ramaswamy}}]{BowickPRX2022}%
  \BibitemOpen
  \bibfield  {author} {\bibinfo {author} {\bibfnamefont {M.~J.}\ \bibnamefont {Bowick}}, \bibinfo {author} {\bibfnamefont {N.}~\bibnamefont {Fakhri}}, \bibinfo {author} {\bibfnamefont {M.~C.}\ \bibnamefont {Marchetti}},\ and\ \bibinfo {author} {\bibfnamefont {S.}~\bibnamefont {Ramaswamy}},\ }\bibfield  {title} {\bibinfo {title} {{Symmetry, Thermodynamics, and Topology in Active Matter}},\ }\href {https://doi.org/10.1103/physrevx.12.010501} {\bibfield  {journal} {\bibinfo  {journal} {Phys. Rev. X}\ }\textbf {\bibinfo {volume} {12}},\ \bibinfo {pages} {010501} (\bibinfo {year} {2022})}\BibitemShut {NoStop}%
\bibitem [{\citenamefont {te~Vrugt}\ and\ \citenamefont {Wittkowski}(2025)}]{teVrugtEPJE2025}%
  \BibitemOpen
  \bibfield  {author} {\bibinfo {author} {\bibfnamefont {M.}~\bibnamefont {te~Vrugt}}\ and\ \bibinfo {author} {\bibfnamefont {R.}~\bibnamefont {Wittkowski}},\ }\bibfield  {title} {\bibinfo {title} {{Metareview: a survey of active matter reviews}},\ }\href {https://doi.org/10.1140/epje/s10189-024-00466-z} {\bibfield  {journal} {\bibinfo  {journal} {Eur. Phys. J. E}\ }\textbf {\bibinfo {volume} {48}},\ \bibinfo {pages} {12} (\bibinfo {year} {2025})}\BibitemShut {NoStop}%
\bibitem [{\citenamefont {J\"{u}licher}\ \emph {et~al.}(1997)\citenamefont {J\"{u}licher}, \citenamefont {Ajdari},\ and\ \citenamefont {Prost}}]{JulicherRMP1997}%
  \BibitemOpen
  \bibfield  {author} {\bibinfo {author} {\bibfnamefont {F.}~\bibnamefont {J\"{u}licher}}, \bibinfo {author} {\bibfnamefont {A.}~\bibnamefont {Ajdari}},\ and\ \bibinfo {author} {\bibfnamefont {J.}~\bibnamefont {Prost}},\ }\bibfield  {title} {\bibinfo {title} {{Modeling molecular motors}},\ }\href {https://doi.org/10.1103/RevModPhys.69.1269} {\bibfield  {journal} {\bibinfo  {journal} {Rev. Mod. Phys.}\ }\textbf {\bibinfo {volume} {69}},\ \bibinfo {pages} {1269} (\bibinfo {year} {1997})}\BibitemShut {NoStop}%
\bibitem [{\citenamefont {Elgeti}\ \emph {et~al.}(2015)\citenamefont {Elgeti}, \citenamefont {Winkler},\ and\ \citenamefont {Gompper}}]{ElgetiRPP2015}%
  \BibitemOpen
  \bibfield  {author} {\bibinfo {author} {\bibfnamefont {J.}~\bibnamefont {Elgeti}}, \bibinfo {author} {\bibfnamefont {R.~G.}\ \bibnamefont {Winkler}},\ and\ \bibinfo {author} {\bibfnamefont {G.}~\bibnamefont {Gompper}},\ }\bibfield  {title} {\bibinfo {title} {{Physics of microswimmers—single particle motion and collective behavior: a review}},\ }\href {https://doi.org/10.1088/0034-4885/78/5/056601} {\bibfield  {journal} {\bibinfo  {journal} {Rep. Prog. Phys.}\ }\textbf {\bibinfo {volume} {78}},\ \bibinfo {pages} {056601} (\bibinfo {year} {2015})}\BibitemShut {NoStop}%
\bibitem [{\citenamefont {Alert}\ and\ \citenamefont {Trepat}(2020)}]{AlertARCMP2020}%
  \BibitemOpen
  \bibfield  {author} {\bibinfo {author} {\bibfnamefont {R.}~\bibnamefont {Alert}}\ and\ \bibinfo {author} {\bibfnamefont {X.}~\bibnamefont {Trepat}},\ }\bibfield  {title} {\bibinfo {title} {{Physical Models of Collective Cell Migration}},\ }\href {https://doi.org/https://doi.org/10.1146/annurev-conmatphys-031218-013516} {\bibfield  {journal} {\bibinfo  {journal} {Annu. Rev. Condens. Matter Phys.}\ }\textbf {\bibinfo {volume} {11}},\ \bibinfo {pages} {77} (\bibinfo {year} {2020})}\BibitemShut {NoStop}%
\bibitem [{\citenamefont {Kumar}\ \emph {et~al.}(2014)\citenamefont {Kumar}, \citenamefont {Soni}, \citenamefont {Ramaswamy},\ and\ \citenamefont {Sood}}]{KumarNatComms2014}%
  \BibitemOpen
  \bibfield  {author} {\bibinfo {author} {\bibfnamefont {N.}~\bibnamefont {Kumar}}, \bibinfo {author} {\bibfnamefont {H.}~\bibnamefont {Soni}}, \bibinfo {author} {\bibfnamefont {S.}~\bibnamefont {Ramaswamy}},\ and\ \bibinfo {author} {\bibfnamefont {A.~K.}\ \bibnamefont {Sood}},\ }\bibfield  {title} {\bibinfo {title} {{Flocking at a distance in active granular matter}},\ }\href {https://doi.org/10.1038/ncomms5688} {\bibfield  {journal} {\bibinfo  {journal} {Nat. Commun.}\ }\textbf {\bibinfo {volume} {5}},\ \bibinfo {pages} {ncomms5688} (\bibinfo {year} {2014})}\BibitemShut {NoStop}%
\bibitem [{\citenamefont {Ouellette}(2022)}]{OuellettePB2022}%
  \BibitemOpen
  \bibfield  {author} {\bibinfo {author} {\bibfnamefont {N.~T.}\ \bibnamefont {Ouellette}},\ }\bibfield  {title} {\bibinfo {title} {{A physics perspective on collective animal behavior}},\ }\href {https://doi.org/10.1088/1478-3975/ac4bef} {\bibfield  {journal} {\bibinfo  {journal} {Phys. Biol.}\ }\textbf {\bibinfo {volume} {19}},\ \bibinfo {pages} {021004} (\bibinfo {year} {2022})}\BibitemShut {NoStop}%
\bibitem [{\citenamefont {Corbetta}\ and\ \citenamefont {Toschi}(2023)}]{CorbettaARCMP2023}%
  \BibitemOpen
  \bibfield  {author} {\bibinfo {author} {\bibfnamefont {A.}~\bibnamefont {Corbetta}}\ and\ \bibinfo {author} {\bibfnamefont {F.}~\bibnamefont {Toschi}},\ }\bibfield  {title} {\bibinfo {title} {{Physics of Human Crowds}},\ }\href {https://doi.org/https://doi.org/10.1146/annurev-conmatphys-031620-100450} {\bibfield  {journal} {\bibinfo  {journal} {Annu. Rev. Condens. Matter Phys.}\ }\textbf {\bibinfo {volume} {14}},\ \bibinfo {pages} {311} (\bibinfo {year} {2023})}\BibitemShut {NoStop}%
\bibitem [{\citenamefont {Galajda}\ \emph {et~al.}(2007)\citenamefont {Galajda}, \citenamefont {Keymer}, \citenamefont {Chaikin},\ and\ \citenamefont {Austin}}]{GalajdaJB2007}%
  \BibitemOpen
  \bibfield  {author} {\bibinfo {author} {\bibfnamefont {P.}~\bibnamefont {Galajda}}, \bibinfo {author} {\bibfnamefont {J.}~\bibnamefont {Keymer}}, \bibinfo {author} {\bibfnamefont {P.}~\bibnamefont {Chaikin}},\ and\ \bibinfo {author} {\bibfnamefont {R.}~\bibnamefont {Austin}},\ }\bibfield  {title} {\bibinfo {title} {{A Wall of Funnels Concentrates Swimming Bacteria}},\ }\href {https://doi.org/10.1128/JB.01033-07} {\bibfield  {journal} {\bibinfo  {journal} {J. Bacteriol.}\ }\textbf {\bibinfo {volume} {189}},\ \bibinfo {pages} {8704} (\bibinfo {year} {2007})}\BibitemShut {NoStop}%
\bibitem [{\citenamefont {Wan}\ \emph {et~al.}(2008)\citenamefont {Wan}, \citenamefont {Olson~Reichhardt}, \citenamefont {Nussinov},\ and\ \citenamefont {Reichhardt}}]{WanPRL2008}%
  \BibitemOpen
  \bibfield  {author} {\bibinfo {author} {\bibfnamefont {M.~B.}\ \bibnamefont {Wan}}, \bibinfo {author} {\bibfnamefont {C.~J.}\ \bibnamefont {Olson~Reichhardt}}, \bibinfo {author} {\bibfnamefont {Z.}~\bibnamefont {Nussinov}},\ and\ \bibinfo {author} {\bibfnamefont {C.}~\bibnamefont {Reichhardt}},\ }\bibfield  {title} {\bibinfo {title} {{Rectification of Swimming Bacteria and Self-Driven Particle Systems by Arrays of Asymmetric Barriers}},\ }\href {https://doi.org/10.1103/PhysRevLett.101.018102} {\bibfield  {journal} {\bibinfo  {journal} {Phys. Rev. Lett.}\ }\textbf {\bibinfo {volume} {101}},\ \bibinfo {pages} {018102} (\bibinfo {year} {2008})}\BibitemShut {NoStop}%
\bibitem [{\citenamefont {Tailleur}\ and\ \citenamefont {Cates}(2009)}]{TailleurEPL2009}%
  \BibitemOpen
  \bibfield  {author} {\bibinfo {author} {\bibfnamefont {J.}~\bibnamefont {Tailleur}}\ and\ \bibinfo {author} {\bibfnamefont {M.~E.}\ \bibnamefont {Cates}},\ }\bibfield  {title} {\bibinfo {title} {{Sedimentation, trapping, and rectification of dilute bacteria}},\ }\href {https://doi.org/10.1209/0295-5075/86/60002} {\bibfield  {journal} {\bibinfo  {journal} {Europhys. Lett.}\ }\textbf {\bibinfo {volume} {86}},\ \bibinfo {pages} {60002} (\bibinfo {year} {2009})}\BibitemShut {NoStop}%
\bibitem [{\citenamefont {Angelani}\ \emph {et~al.}(2009)\citenamefont {Angelani}, \citenamefont {Di~Leonardo},\ and\ \citenamefont {Ruocco}}]{AngelaniPRL2009}%
  \BibitemOpen
  \bibfield  {author} {\bibinfo {author} {\bibfnamefont {L.}~\bibnamefont {Angelani}}, \bibinfo {author} {\bibfnamefont {R.}~\bibnamefont {Di~Leonardo}},\ and\ \bibinfo {author} {\bibfnamefont {G.}~\bibnamefont {Ruocco}},\ }\bibfield  {title} {\bibinfo {title} {{Self-Starting Micromotors in a Bacterial Bath}},\ }\href {https://doi.org/10.1103/PhysRevLett.102.048104} {\bibfield  {journal} {\bibinfo  {journal} {Phys. Rev. Lett.}\ }\textbf {\bibinfo {volume} {102}},\ \bibinfo {pages} {048104} (\bibinfo {year} {2009})}\BibitemShut {NoStop}%
\bibitem [{\citenamefont {Angelani}\ and\ \citenamefont {Leonardo}(2010)}]{AngelaniNJP2010}%
  \BibitemOpen
  \bibfield  {author} {\bibinfo {author} {\bibfnamefont {L.}~\bibnamefont {Angelani}}\ and\ \bibinfo {author} {\bibfnamefont {R.~D.}\ \bibnamefont {Leonardo}},\ }\bibfield  {title} {\bibinfo {title} {{Geometrically biased random walks in bacteria-driven micro-shuttles}},\ }\href {https://doi.org/10.1088/1367-2630/12/11/113017} {\bibfield  {journal} {\bibinfo  {journal} {New J. Phys.}\ }\textbf {\bibinfo {volume} {12}},\ \bibinfo {pages} {113017} (\bibinfo {year} {2010})}\BibitemShut {NoStop}%
\bibitem [{\citenamefont {Sokolov}\ \emph {et~al.}(2010)\citenamefont {Sokolov}, \citenamefont {Apodaca}, \citenamefont {Grzybowski},\ and\ \citenamefont {Aranson}}]{SokolovPNAS2010}%
  \BibitemOpen
  \bibfield  {author} {\bibinfo {author} {\bibfnamefont {A.}~\bibnamefont {Sokolov}}, \bibinfo {author} {\bibfnamefont {M.~M.}\ \bibnamefont {Apodaca}}, \bibinfo {author} {\bibfnamefont {B.~A.}\ \bibnamefont {Grzybowski}},\ and\ \bibinfo {author} {\bibfnamefont {I.~S.}\ \bibnamefont {Aranson}},\ }\bibfield  {title} {\bibinfo {title} {{Swimming bacteria power microscopic gears}},\ }\href {https://doi.org/10.1073/pnas.0913015107} {\bibfield  {journal} {\bibinfo  {journal} {Proc. Natl. Acad. Sci.}\ }\textbf {\bibinfo {volume} {107}},\ \bibinfo {pages} {969} (\bibinfo {year} {2010})}\BibitemShut {NoStop}%
\bibitem [{\citenamefont {{Di Leonardo}}\ \emph {et~al.}(2010)\citenamefont {{Di Leonardo}}, \citenamefont {Angelani}, \citenamefont {Dell'Arciprete}, \citenamefont {Ruocco}, \citenamefont {Iebba}, \citenamefont {Schippa}, \citenamefont {Conte}, \citenamefont {Mecarini}, \citenamefont {Angelis},\ and\ \citenamefont {Fabrizio}}]{DiLeonardoPNAS2010}%
  \BibitemOpen
  \bibfield  {author} {\bibinfo {author} {\bibfnamefont {R.}~\bibnamefont {{Di Leonardo}}}, \bibinfo {author} {\bibfnamefont {L.}~\bibnamefont {Angelani}}, \bibinfo {author} {\bibfnamefont {D.}~\bibnamefont {Dell'Arciprete}}, \bibinfo {author} {\bibfnamefont {G.}~\bibnamefont {Ruocco}}, \bibinfo {author} {\bibfnamefont {V.}~\bibnamefont {Iebba}}, \bibinfo {author} {\bibfnamefont {S.}~\bibnamefont {Schippa}}, \bibinfo {author} {\bibfnamefont {M.~P.}\ \bibnamefont {Conte}}, \bibinfo {author} {\bibfnamefont {F.}~\bibnamefont {Mecarini}}, \bibinfo {author} {\bibfnamefont {F.~D.}\ \bibnamefont {Angelis}},\ and\ \bibinfo {author} {\bibfnamefont {E.~D.}\ \bibnamefont {Fabrizio}},\ }\bibfield  {title} {\bibinfo {title} {{Bacterial ratchet motors}},\ }\href {https://doi.org/10.1073/pnas.0910426107} {\bibfield  {journal} {\bibinfo  {journal} {Proc. Natl. Acad. Sci.}\ }\textbf {\bibinfo {volume} {107}},\ \bibinfo {pages} {9541} (\bibinfo {year} {2010})}\BibitemShut {NoStop}%
\bibitem [{\citenamefont {Kaiser}\ \emph {et~al.}(2014)\citenamefont {Kaiser}, \citenamefont {Peshkov}, \citenamefont {Sokolov}, \citenamefont {ten Hagen}, \citenamefont {L\"owen},\ and\ \citenamefont {Aranson}}]{KaiserPRL2014}%
  \BibitemOpen
  \bibfield  {author} {\bibinfo {author} {\bibfnamefont {A.}~\bibnamefont {Kaiser}}, \bibinfo {author} {\bibfnamefont {A.}~\bibnamefont {Peshkov}}, \bibinfo {author} {\bibfnamefont {A.}~\bibnamefont {Sokolov}}, \bibinfo {author} {\bibfnamefont {B.}~\bibnamefont {ten Hagen}}, \bibinfo {author} {\bibfnamefont {H.}~\bibnamefont {L\"owen}},\ and\ \bibinfo {author} {\bibfnamefont {I.~S.}\ \bibnamefont {Aranson}},\ }\bibfield  {title} {\bibinfo {title} {{Transport Powered by Bacterial Turbulence}},\ }\href {https://doi.org/10.1103/PhysRevLett.112.158101} {\bibfield  {journal} {\bibinfo  {journal} {Phys. Rev. Lett.}\ }\textbf {\bibinfo {volume} {112}},\ \bibinfo {pages} {158101} (\bibinfo {year} {2014})}\BibitemShut {NoStop}%
\bibitem [{\citenamefont {Mallory}\ \emph {et~al.}(2014)\citenamefont {Mallory}, \citenamefont {Valeriani},\ and\ \citenamefont {Cacciuto}}]{MalloryPRE2014}%
  \BibitemOpen
  \bibfield  {author} {\bibinfo {author} {\bibfnamefont {S.~A.}\ \bibnamefont {Mallory}}, \bibinfo {author} {\bibfnamefont {C.}~\bibnamefont {Valeriani}},\ and\ \bibinfo {author} {\bibfnamefont {A.}~\bibnamefont {Cacciuto}},\ }\bibfield  {title} {\bibinfo {title} {{Curvature-induced activation of a passive tracer in an active bath}},\ }\href {https://doi.org/10.1103/PhysRevE.90.032309} {\bibfield  {journal} {\bibinfo  {journal} {Phys. Rev. E}\ }\textbf {\bibinfo {volume} {90}},\ \bibinfo {pages} {032309} (\bibinfo {year} {2014})}\BibitemShut {NoStop}%
\bibitem [{\citenamefont {Baek}\ \emph {et~al.}(2018)\citenamefont {Baek}, \citenamefont {Solon}, \citenamefont {Xu}, \citenamefont {Nikola},\ and\ \citenamefont {Kafri}}]{BaekPRL2018}%
  \BibitemOpen
  \bibfield  {author} {\bibinfo {author} {\bibfnamefont {Y.}~\bibnamefont {Baek}}, \bibinfo {author} {\bibfnamefont {A.~P.}\ \bibnamefont {Solon}}, \bibinfo {author} {\bibfnamefont {X.}~\bibnamefont {Xu}}, \bibinfo {author} {\bibfnamefont {N.}~\bibnamefont {Nikola}},\ and\ \bibinfo {author} {\bibfnamefont {Y.}~\bibnamefont {Kafri}},\ }\bibfield  {title} {\bibinfo {title} {{Generic Long-Range Interactions Between Passive Bodies in an Active Fluid}},\ }\href {https://doi.org/10.1103/PhysRevLett.120.058002} {\bibfield  {journal} {\bibinfo  {journal} {Phys. Rev. Lett.}\ }\textbf {\bibinfo {volume} {120}},\ \bibinfo {pages} {058002} (\bibinfo {year} {2018})}\BibitemShut {NoStop}%
\bibitem [{\citenamefont {Granek}\ \emph {et~al.}(2020)\citenamefont {Granek}, \citenamefont {Baek}, \citenamefont {Kafri},\ and\ \citenamefont {Solon}}]{GranekJSM2020}%
  \BibitemOpen
  \bibfield  {author} {\bibinfo {author} {\bibfnamefont {O.}~\bibnamefont {Granek}}, \bibinfo {author} {\bibfnamefont {Y.}~\bibnamefont {Baek}}, \bibinfo {author} {\bibfnamefont {Y.}~\bibnamefont {Kafri}},\ and\ \bibinfo {author} {\bibfnamefont {A.~P.}\ \bibnamefont {Solon}},\ }\bibfield  {title} {\bibinfo {title} {{Bodies in an interacting active fluid: far-field influence of a single body and interaction between two bodies}},\ }\href {https://doi.org/10.1088/1742-5468/ab7f34} {\bibfield  {journal} {\bibinfo  {journal} {J. Stat. Mech.: Theor. Exp.}\ }\textbf {\bibinfo {volume} {2020}},\ \bibinfo {pages} {063211} (\bibinfo {year} {2020})}\BibitemShut {NoStop}%
\bibitem [{\citenamefont {Cates}\ and\ \citenamefont {Tailleur}(2015)}]{CatesARCMP2015}%
  \BibitemOpen
  \bibfield  {author} {\bibinfo {author} {\bibfnamefont {M.~E.}\ \bibnamefont {Cates}}\ and\ \bibinfo {author} {\bibfnamefont {J.}~\bibnamefont {Tailleur}},\ }\bibfield  {title} {\bibinfo {title} {{Motility-Induced Phase Separation}},\ }\href {https://doi.org/https://doi.org/10.1146/annurev-conmatphys-031214-014710} {\bibfield  {journal} {\bibinfo  {journal} {Annu. Rev. Condens. Matter Phys.}\ }\textbf {\bibinfo {volume} {6}},\ \bibinfo {pages} {219} (\bibinfo {year} {2015})}\BibitemShut {NoStop}%
\bibitem [{\citenamefont {Bialk\'{e}}\ \emph {et~al.}(2015)\citenamefont {Bialk\'{e}}, \citenamefont {Speck},\ and\ \citenamefont {L\"{o}wen}}]{BialkeJNCS2015}%
  \BibitemOpen
  \bibfield  {author} {\bibinfo {author} {\bibfnamefont {J.}~\bibnamefont {Bialk\'{e}}}, \bibinfo {author} {\bibfnamefont {T.}~\bibnamefont {Speck}},\ and\ \bibinfo {author} {\bibfnamefont {H.}~\bibnamefont {L\"{o}wen}},\ }\bibfield  {title} {\bibinfo {title} {{Active colloidal suspensions: Clustering and phase behavior}},\ }\href {https://doi.org/https://doi.org/10.1016/j.jnoncrysol.2014.08.011} {\bibfield  {journal} {\bibinfo  {journal} {J. Non-Cryst. Solids}\ }\textbf {\bibinfo {volume} {407}},\ \bibinfo {pages} {367} (\bibinfo {year} {2015})},\ \bibinfo {note} {7th IDMRCS: Relaxation in Complex Systems}\BibitemShut {NoStop}%
\bibitem [{\citenamefont {O'Byrne}\ \emph {et~al.}(2023)\citenamefont {O'Byrne}, \citenamefont {Solon}, \citenamefont {Tailleur},\ and\ \citenamefont {Zhao}}]{O'ByrneBook2023}%
  \BibitemOpen
  \bibfield  {author} {\bibinfo {author} {\bibfnamefont {J.}~\bibnamefont {O'Byrne}}, \bibinfo {author} {\bibfnamefont {A.}~\bibnamefont {Solon}}, \bibinfo {author} {\bibfnamefont {J.}~\bibnamefont {Tailleur}},\ and\ \bibinfo {author} {\bibfnamefont {Y.}~\bibnamefont {Zhao}},\ }\bibfield  {title} {\bibinfo {title} {{An Introduction to Motility-induced Phase Separation}},\ }in\ \href {https://doi.org/10.1039/9781839169465-00107} {\emph {\bibinfo {booktitle} {{Out-of-equilibrium Soft Matter}}}}\ (\bibinfo  {publisher} {The Royal Society of Chemistry},\ \bibinfo {year} {2023})\BibitemShut {NoStop}%
\bibitem [{\citenamefont {Toner}\ \emph {et~al.}(2005)\citenamefont {Toner}, \citenamefont {Tu},\ and\ \citenamefont {Ramaswamy}}]{TonerAP2005}%
  \BibitemOpen
  \bibfield  {author} {\bibinfo {author} {\bibfnamefont {J.}~\bibnamefont {Toner}}, \bibinfo {author} {\bibfnamefont {Y.}~\bibnamefont {Tu}},\ and\ \bibinfo {author} {\bibfnamefont {S.}~\bibnamefont {Ramaswamy}},\ }\bibfield  {title} {\bibinfo {title} {{Hydrodynamics and phases of flocks}},\ }\href {https://doi.org/https://doi.org/10.1016/j.aop.2005.04.011} {\bibfield  {journal} {\bibinfo  {journal} {Annal. Phys.}\ }\textbf {\bibinfo {volume} {318}},\ \bibinfo {pages} {170} (\bibinfo {year} {2005})},\ \bibinfo {note} {special Issue}\BibitemShut {NoStop}%
\bibitem [{\citenamefont {Vicsek}\ and\ \citenamefont {Zafeiris}(2012)}]{VicsekPR2012}%
  \BibitemOpen
  \bibfield  {author} {\bibinfo {author} {\bibfnamefont {T.}~\bibnamefont {Vicsek}}\ and\ \bibinfo {author} {\bibfnamefont {A.}~\bibnamefont {Zafeiris}},\ }\bibfield  {title} {\bibinfo {title} {{Collective motion}},\ }\href {https://doi.org/https://doi.org/10.1016/j.physrep.2012.03.004} {\bibfield  {journal} {\bibinfo  {journal} {Phys. Rep.}\ }\textbf {\bibinfo {volume} {517}},\ \bibinfo {pages} {71} (\bibinfo {year} {2012})},\ \bibinfo {note} {collective motion}\BibitemShut {NoStop}%
\bibitem [{\citenamefont {Cavagna}\ and\ \citenamefont {Giardina}(2014)}]{CavagnaARCMP2014}%
  \BibitemOpen
  \bibfield  {author} {\bibinfo {author} {\bibfnamefont {A.}~\bibnamefont {Cavagna}}\ and\ \bibinfo {author} {\bibfnamefont {I.}~\bibnamefont {Giardina}},\ }\bibfield  {title} {\bibinfo {title} {Bird flocks as condensed matter},\ }\href {https://doi.org/https://doi.org/10.1146/annurev-conmatphys-031113-133834} {\bibfield  {journal} {\bibinfo  {journal} {Annu. Rev. Condens. Matter Phys.}\ }\textbf {\bibinfo {volume} {5}},\ \bibinfo {pages} {183} (\bibinfo {year} {2014})}\BibitemShut {NoStop}%
\bibitem [{\citenamefont {Ginelli}(2016)}]{GinelliEPJST2016}%
  \BibitemOpen
  \bibfield  {author} {\bibinfo {author} {\bibfnamefont {F.}~\bibnamefont {Ginelli}},\ }\bibfield  {title} {\bibinfo {title} {{The Physics of the Vicsek model}},\ }\href {https://doi.org/10.1140/epjst/e2016-60066-8} {\bibfield  {journal} {\bibinfo  {journal} {Eur. Phys. J. Spec. Top.}\ }\textbf {\bibinfo {volume} {225}},\ \bibinfo {pages} {2099} (\bibinfo {year} {2016})}\BibitemShut {NoStop}%
\bibitem [{\citenamefont {Chat\'{e}}(2020)}]{ChateARCMP2020}%
  \BibitemOpen
  \bibfield  {author} {\bibinfo {author} {\bibfnamefont {H.}~\bibnamefont {Chat\'{e}}},\ }\bibfield  {title} {\bibinfo {title} {{Dry Aligning Dilute Active Matter}},\ }\href {https://doi.org/https://doi.org/10.1146/annurev-conmatphys-031119-050752} {\bibfield  {journal} {\bibinfo  {journal} {Annu. Rev. Condens. Matter Phys.}\ }\textbf {\bibinfo {volume} {11}},\ \bibinfo {pages} {189} (\bibinfo {year} {2020})}\BibitemShut {NoStop}%
\bibitem [{\citenamefont {Fodor}\ \emph {et~al.}(2022)\citenamefont {Fodor}, \citenamefont {Jack},\ and\ \citenamefont {Cates}}]{FodorARCMP2022}%
  \BibitemOpen
  \bibfield  {author} {\bibinfo {author} {\bibfnamefont {{\'{E}}.}~\bibnamefont {Fodor}}, \bibinfo {author} {\bibfnamefont {R.~L.}\ \bibnamefont {Jack}},\ and\ \bibinfo {author} {\bibfnamefont {M.~E.}\ \bibnamefont {Cates}},\ }\bibfield  {title} {\bibinfo {title} {{Irreversibility and Biased Ensembles in Active Matter: Insights from Stochastic Thermodynamics}},\ }\href {https://doi.org/https://doi.org/10.1146/annurev-conmatphys-031720-032419} {\bibfield  {journal} {\bibinfo  {journal} {Annu. Rev. Condens. Matter Phys.}\ }\textbf {\bibinfo {volume} {13}},\ \bibinfo {pages} {215} (\bibinfo {year} {2022})}\BibitemShut {NoStop}%
\bibitem [{\citenamefont {Nardini}\ \emph {et~al.}(2017)\citenamefont {Nardini}, \citenamefont {Fodor}, \citenamefont {Tjhung}, \citenamefont {van Wijland}, \citenamefont {Tailleur},\ and\ \citenamefont {Cates}}]{NardiniPRX2017}%
  \BibitemOpen
  \bibfield  {author} {\bibinfo {author} {\bibfnamefont {C.}~\bibnamefont {Nardini}}, \bibinfo {author} {\bibfnamefont {{\'{E}}.}~\bibnamefont {Fodor}}, \bibinfo {author} {\bibfnamefont {E.}~\bibnamefont {Tjhung}}, \bibinfo {author} {\bibfnamefont {F.}~\bibnamefont {van Wijland}}, \bibinfo {author} {\bibfnamefont {J.}~\bibnamefont {Tailleur}},\ and\ \bibinfo {author} {\bibfnamefont {M.~E.}\ \bibnamefont {Cates}},\ }\bibfield  {title} {\bibinfo {title} {{Entropy Production in Field Theories without Time-Reversal Symmetry: Quantifying the Non-Equilibrium Character of Active Matter}},\ }\href {https://doi.org/10.1103/PhysRevX.7.021007} {\bibfield  {journal} {\bibinfo  {journal} {Phys. Rev. X}\ }\textbf {\bibinfo {volume} {7}},\ \bibinfo {pages} {021007} (\bibinfo {year} {2017})}\BibitemShut {NoStop}%
\bibitem [{\citenamefont {Crosato}\ \emph {et~al.}(2019)\citenamefont {Crosato}, \citenamefont {Prokopenko},\ and\ \citenamefont {Spinney}}]{CrosatoPRE2019}%
  \BibitemOpen
  \bibfield  {author} {\bibinfo {author} {\bibfnamefont {E.}~\bibnamefont {Crosato}}, \bibinfo {author} {\bibfnamefont {M.}~\bibnamefont {Prokopenko}},\ and\ \bibinfo {author} {\bibfnamefont {R.~E.}\ \bibnamefont {Spinney}},\ }\bibfield  {title} {\bibinfo {title} {{Irreversibility and emergent structure in active matter}},\ }\href {https://journals.aps.org/pre/abstract/10.1103/PhysRevE.100.042613} {\bibfield  {journal} {\bibinfo  {journal} {Phys. Rev. E}\ }\textbf {\bibinfo {volume} {100}},\ \bibinfo {pages} {042613} (\bibinfo {year} {2019})}\BibitemShut {NoStop}%
\bibitem [{\citenamefont {O’Byrne}\ \emph {et~al.}(2022)\citenamefont {O’Byrne}, \citenamefont {Kafri}, \citenamefont {Tailleur},\ and\ \citenamefont {Wijland}}]{O'ByrneNRP2022}%
  \BibitemOpen
  \bibfield  {author} {\bibinfo {author} {\bibfnamefont {J.}~\bibnamefont {O’Byrne}}, \bibinfo {author} {\bibfnamefont {Y.}~\bibnamefont {Kafri}}, \bibinfo {author} {\bibfnamefont {J.}~\bibnamefont {Tailleur}},\ and\ \bibinfo {author} {\bibfnamefont {F.~v.}\ \bibnamefont {Wijland}},\ }\bibfield  {title} {\bibinfo {title} {{Time irreversibility in active matter, from micro to macro}},\ }\href {https://doi.org/10.1038/s42254-021-00406-2} {\bibfield  {journal} {\bibinfo  {journal} {Nat. Rev. Phys.}\ }\textbf {\bibinfo {volume} {4}},\ \bibinfo {pages} {167} (\bibinfo {year} {2022})}\BibitemShut {NoStop}%
\bibitem [{\citenamefont {Ro}\ \emph {et~al.}(2022)\citenamefont {Ro}, \citenamefont {Guo}, \citenamefont {Shih}, \citenamefont {Phan}, \citenamefont {Austin}, \citenamefont {Levine}, \citenamefont {Chaikin},\ and\ \citenamefont {Martiniani}}]{RoPRL2022}%
  \BibitemOpen
  \bibfield  {author} {\bibinfo {author} {\bibfnamefont {S.}~\bibnamefont {Ro}}, \bibinfo {author} {\bibfnamefont {B.}~\bibnamefont {Guo}}, \bibinfo {author} {\bibfnamefont {A.}~\bibnamefont {Shih}}, \bibinfo {author} {\bibfnamefont {T.~V.}\ \bibnamefont {Phan}}, \bibinfo {author} {\bibfnamefont {R.~H.}\ \bibnamefont {Austin}}, \bibinfo {author} {\bibfnamefont {D.}~\bibnamefont {Levine}}, \bibinfo {author} {\bibfnamefont {P.~M.}\ \bibnamefont {Chaikin}},\ and\ \bibinfo {author} {\bibfnamefont {S.}~\bibnamefont {Martiniani}},\ }\bibfield  {title} {\bibinfo {title} {{Model-Free Measurement of Local Entropy Production and Extractable Work in Active Matter}},\ }\href {https://doi.org/10.1103/PhysRevLett.129.220601} {\bibfield  {journal} {\bibinfo  {journal} {Phys. Rev. Lett.}\ }\textbf {\bibinfo {volume} {129}},\ \bibinfo {pages} {220601} (\bibinfo {year} {2022})}\BibitemShut {NoStop}%
\bibitem [{\citenamefont {Pietzonka}\ and\ \citenamefont {Seifert}(2017)}]{PietzonkaJPA2018}%
  \BibitemOpen
  \bibfield  {author} {\bibinfo {author} {\bibfnamefont {P.}~\bibnamefont {Pietzonka}}\ and\ \bibinfo {author} {\bibfnamefont {U.}~\bibnamefont {Seifert}},\ }\bibfield  {title} {\bibinfo {title} {{Entropy production of active particles and for particles in active baths}},\ }\href {https://doi.org/10.1088/1751-8121/aa91b9} {\bibfield  {journal} {\bibinfo  {journal} {J. Phys. A: Math. Theor.}\ }\textbf {\bibinfo {volume} {51}},\ \bibinfo {pages} {01LT01} (\bibinfo {year} {2017})}\BibitemShut {NoStop}%
\bibitem [{\citenamefont {Dadhichi}\ \emph {et~al.}(2018)\citenamefont {Dadhichi}, \citenamefont {Maitra},\ and\ \citenamefont {Ramaswamy}}]{DadhichiJSM2018}%
  \BibitemOpen
  \bibfield  {author} {\bibinfo {author} {\bibfnamefont {L.~P.}\ \bibnamefont {Dadhichi}}, \bibinfo {author} {\bibfnamefont {A.}~\bibnamefont {Maitra}},\ and\ \bibinfo {author} {\bibfnamefont {S.}~\bibnamefont {Ramaswamy}},\ }\bibfield  {title} {\bibinfo {title} {{Origins and diagnostics of the nonequilibrium character of active systems}},\ }\href {https://iopscience.iop.org/article/10.1088/1742-5468/aae852/meta} {\bibfield  {journal} {\bibinfo  {journal} {J. Stat. Mech.: Theor. Expr.}\ }\textbf {\bibinfo {volume} {2018}},\ \bibinfo {pages} {123201} (\bibinfo {year} {2018})}\BibitemShut {NoStop}%
\bibitem [{\citenamefont {Speck}(2018)}]{SpeckEPL2018}%
  \BibitemOpen
  \bibfield  {author} {\bibinfo {author} {\bibfnamefont {T.}~\bibnamefont {Speck}},\ }\bibfield  {title} {\bibinfo {title} {{Active Brownian particles driven by constant affinity}},\ }\href {https://doi.org/10.1209/0295-5075/123/20007} {\bibfield  {journal} {\bibinfo  {journal} {Europhys. Lett.}\ }\textbf {\bibinfo {volume} {123}},\ \bibinfo {pages} {20007} (\bibinfo {year} {2018})}\BibitemShut {NoStop}%
\bibitem [{\citenamefont {Speck}(2019)}]{SpeckPRE2019}%
  \BibitemOpen
  \bibfield  {author} {\bibinfo {author} {\bibfnamefont {T.}~\bibnamefont {Speck}},\ }\bibfield  {title} {\bibinfo {title} {{Thermodynamic approach to the self-diffusiophoresis of colloidal Janus particles}},\ }\href {https://doi.org/10.1103/PhysRevE.99.060602} {\bibfield  {journal} {\bibinfo  {journal} {Phys. Rev. E}\ }\textbf {\bibinfo {volume} {99}},\ \bibinfo {pages} {060602} (\bibinfo {year} {2019})}\BibitemShut {NoStop}%
\bibitem [{\citenamefont {Dabelow}\ \emph {et~al.}(2019)\citenamefont {Dabelow}, \citenamefont {Bo},\ and\ \citenamefont {Eichhorn}}]{DabelowPRX2019}%
  \BibitemOpen
  \bibfield  {author} {\bibinfo {author} {\bibfnamefont {L.}~\bibnamefont {Dabelow}}, \bibinfo {author} {\bibfnamefont {S.}~\bibnamefont {Bo}},\ and\ \bibinfo {author} {\bibfnamefont {R.}~\bibnamefont {Eichhorn}},\ }\bibfield  {title} {\bibinfo {title} {{Irreversibility in active matter systems: Fluctuation theorem and mutual information}},\ }\href {https://journals.aps.org/prx/abstract/10.1103/PhysRevX.9.021009} {\bibfield  {journal} {\bibinfo  {journal} {Phys. Rev. X}\ }\textbf {\bibinfo {volume} {9}},\ \bibinfo {pages} {021009} (\bibinfo {year} {2019})}\BibitemShut {NoStop}%
\bibitem [{\citenamefont {Markovich}\ \emph {et~al.}(2021)\citenamefont {Markovich}, \citenamefont {Fodor}, \citenamefont {Tjhung},\ and\ \citenamefont {Cates}}]{MarkovichPRX2021}%
  \BibitemOpen
  \bibfield  {author} {\bibinfo {author} {\bibfnamefont {T.}~\bibnamefont {Markovich}}, \bibinfo {author} {\bibfnamefont {{\'{E}}.}~\bibnamefont {Fodor}}, \bibinfo {author} {\bibfnamefont {E.}~\bibnamefont {Tjhung}},\ and\ \bibinfo {author} {\bibfnamefont {M.~E.}\ \bibnamefont {Cates}},\ }\bibfield  {title} {\bibinfo {title} {{Thermodynamics of Active Field Theories: Energetic Cost of Coupling to Reservoirs}},\ }\href {https://doi.org/10.1103/PhysRevX.11.021057} {\bibfield  {journal} {\bibinfo  {journal} {Phys. Rev. X}\ }\textbf {\bibinfo {volume} {11}},\ \bibinfo {pages} {021057} (\bibinfo {year} {2021})}\BibitemShut {NoStop}%
\bibitem [{\citenamefont {Oh}\ and\ \citenamefont {Baek}(2023)}]{OhPRE2023}%
  \BibitemOpen
  \bibfield  {author} {\bibinfo {author} {\bibfnamefont {Y.}~\bibnamefont {Oh}}\ and\ \bibinfo {author} {\bibfnamefont {Y.}~\bibnamefont {Baek}},\ }\bibfield  {title} {\bibinfo {title} {{Effects of the self-propulsion parity on the efficiency of a fuel-consuming active heat engine}},\ }\href {https://doi.org/10.1103/PhysRevE.108.024602} {\bibfield  {journal} {\bibinfo  {journal} {Phys. Rev. E}\ }\textbf {\bibinfo {volume} {108}},\ \bibinfo {pages} {024602} (\bibinfo {year} {2023})}\BibitemShut {NoStop}%
\bibitem [{\citenamefont {Bebon}\ \emph {et~al.}(2024)\citenamefont {Bebon}, \citenamefont {Robinson},\ and\ \citenamefont {Speck}}]{Bebon2024}%
  \BibitemOpen
  \bibfield  {author} {\bibinfo {author} {\bibfnamefont {R.}~\bibnamefont {Bebon}}, \bibinfo {author} {\bibfnamefont {J.~F.}\ \bibnamefont {Robinson}},\ and\ \bibinfo {author} {\bibfnamefont {T.}~\bibnamefont {Speck}},\ }\href {https://arxiv.org/abs/2401.02252} {\bibinfo {title} {{Thermodynamics of active matter: Tracking dissipation across scales}}} (\bibinfo {year} {2024}),\ \Eprint {https://arxiv.org/abs/2401.02252} {arXiv:2401.02252 [cond-mat.soft]} \BibitemShut {NoStop}%
\bibitem [{\citenamefont {Levis}\ and\ \citenamefont {Berthier}(2014)}]{LevisPRE2014}%
  \BibitemOpen
  \bibfield  {author} {\bibinfo {author} {\bibfnamefont {D.}~\bibnamefont {Levis}}\ and\ \bibinfo {author} {\bibfnamefont {L.}~\bibnamefont {Berthier}},\ }\bibfield  {title} {\bibinfo {title} {{Clustering and heterogeneous dynamics in a kinetic Monte Carlo model of self-propelled hard disks}},\ }\href {https://doi.org/10.1103/PhysRevE.89.062301} {\bibfield  {journal} {\bibinfo  {journal} {Phys. Rev. E}\ }\textbf {\bibinfo {volume} {89}},\ \bibinfo {pages} {062301} (\bibinfo {year} {2014})}\BibitemShut {NoStop}%
\bibitem [{\citenamefont {Klamser}\ \emph {et~al.}(2018)\citenamefont {Klamser}, \citenamefont {Kapfer},\ and\ \citenamefont {Krauth}}]{KlamserNatComms2018}%
  \BibitemOpen
  \bibfield  {author} {\bibinfo {author} {\bibfnamefont {J.~U.}\ \bibnamefont {Klamser}}, \bibinfo {author} {\bibfnamefont {S.~C.}\ \bibnamefont {Kapfer}},\ and\ \bibinfo {author} {\bibfnamefont {W.}~\bibnamefont {Krauth}},\ }\bibfield  {title} {\bibinfo {title} {{Thermodynamic phases in two-dimensional active matter}},\ }\href {https://doi.org/10.1038/s41467-018-07491-5} {\bibfield  {journal} {\bibinfo  {journal} {Nat. Commun.}\ }\textbf {\bibinfo {volume} {9}},\ \bibinfo {pages} {5045} (\bibinfo {year} {2018})}\BibitemShut {NoStop}%
\bibitem [{\citenamefont {Berthier}(2014)}]{BerthierPRL2014}%
  \BibitemOpen
  \bibfield  {author} {\bibinfo {author} {\bibfnamefont {L.}~\bibnamefont {Berthier}},\ }\bibfield  {title} {\bibinfo {title} {{Nonequilibrium Glassy Dynamics of Self-Propelled Hard Disks}},\ }\href {https://doi.org/10.1103/PhysRevLett.112.220602} {\bibfield  {journal} {\bibinfo  {journal} {Phys. Rev. Lett.}\ }\textbf {\bibinfo {volume} {112}},\ \bibinfo {pages} {220602} (\bibinfo {year} {2014})}\BibitemShut {NoStop}%
\bibitem [{\citenamefont {Klamser}\ \emph {et~al.}(2019)\citenamefont {Klamser}, \citenamefont {Kapfer},\ and\ \citenamefont {Krauth}}]{KlamserJCP2019}%
  \BibitemOpen
  \bibfield  {author} {\bibinfo {author} {\bibfnamefont {J.~U.}\ \bibnamefont {Klamser}}, \bibinfo {author} {\bibfnamefont {S.~C.}\ \bibnamefont {Kapfer}},\ and\ \bibinfo {author} {\bibfnamefont {W.}~\bibnamefont {Krauth}},\ }\bibfield  {title} {\bibinfo {title} {{A kinetic-Monte Carlo perspective on active matter}},\ }\href {https://doi.org/10.1063/1.5085828} {\bibfield  {journal} {\bibinfo  {journal} {J. Chem. Phys.}\ }\textbf {\bibinfo {volume} {150}},\ \bibinfo {pages} {144113} (\bibinfo {year} {2019})}\BibitemShut {NoStop}%
\bibitem [{\citenamefont {Levis}\ and\ \citenamefont {Berthier}(2015)}]{LevisEPL2015}%
  \BibitemOpen
  \bibfield  {author} {\bibinfo {author} {\bibfnamefont {D.}~\bibnamefont {Levis}}\ and\ \bibinfo {author} {\bibfnamefont {L.}~\bibnamefont {Berthier}},\ }\bibfield  {title} {\bibinfo {title} {{From single-particle to collective effective temperatures in an active fluid of self-propelled particles}},\ }\href {https://doi.org/10.1209/0295-5075/111/60006} {\bibfield  {journal} {\bibinfo  {journal} {Europhys. Lett.}\ }\textbf {\bibinfo {volume} {111}},\ \bibinfo {pages} {60006} (\bibinfo {year} {2015})}\BibitemShut {NoStop}%
\bibitem [{\citenamefont {Klamser}\ \emph {et~al.}(2021)\citenamefont {Klamser}, \citenamefont {Dauchot},\ and\ \citenamefont {Tailleur}}]{KlamserPRL2021}%
  \BibitemOpen
  \bibfield  {author} {\bibinfo {author} {\bibfnamefont {J.~U.}\ \bibnamefont {Klamser}}, \bibinfo {author} {\bibfnamefont {O.}~\bibnamefont {Dauchot}},\ and\ \bibinfo {author} {\bibfnamefont {J.}~\bibnamefont {Tailleur}},\ }\bibfield  {title} {\bibinfo {title} {{Kinetic Monte Carlo Algorithms for Active Matter Systems}},\ }\href {https://doi.org/10.1103/PhysRevLett.127.150602} {\bibfield  {journal} {\bibinfo  {journal} {Phys. Rev. Lett.}\ }\textbf {\bibinfo {volume} {127}},\ \bibinfo {pages} {150602} (\bibinfo {year} {2021})}\BibitemShut {NoStop}%
\bibitem [{\citenamefont {Schnitzer}(1993)}]{SchnitzerPRE1993}%
  \BibitemOpen
  \bibfield  {author} {\bibinfo {author} {\bibfnamefont {M.~J.}\ \bibnamefont {Schnitzer}},\ }\bibfield  {title} {\bibinfo {title} {{Theory of continuum random walks and application to chemotaxis}},\ }\href {https://doi.org/10.1103/PhysRevE.48.2553} {\bibfield  {journal} {\bibinfo  {journal} {Phys. Rev. E}\ }\textbf {\bibinfo {volume} {48}},\ \bibinfo {pages} {2553} (\bibinfo {year} {1993})}\BibitemShut {NoStop}%
\bibitem [{\citenamefont {Schweitzer}\ \emph {et~al.}(1998)\citenamefont {Schweitzer}, \citenamefont {Ebeling},\ and\ \citenamefont {Tilch}}]{SchweitzerPRL1998}%
  \BibitemOpen
  \bibfield  {author} {\bibinfo {author} {\bibfnamefont {F.}~\bibnamefont {Schweitzer}}, \bibinfo {author} {\bibfnamefont {W.}~\bibnamefont {Ebeling}},\ and\ \bibinfo {author} {\bibfnamefont {B.}~\bibnamefont {Tilch}},\ }\bibfield  {title} {\bibinfo {title} {{Complex Motion of Brownian Particles with Energy Depots}},\ }\href {https://doi.org/10.1103/PhysRevLett.80.5044} {\bibfield  {journal} {\bibinfo  {journal} {Phys. Rev. Lett.}\ }\textbf {\bibinfo {volume} {80}},\ \bibinfo {pages} {5044} (\bibinfo {year} {1998})}\BibitemShut {NoStop}%
\bibitem [{\citenamefont {Szamel}(2014)}]{SzamelPRE2014}%
  \BibitemOpen
  \bibfield  {author} {\bibinfo {author} {\bibfnamefont {G.}~\bibnamefont {Szamel}},\ }\bibfield  {title} {\bibinfo {title} {{Self-propelled particle in an external potential: Existence of an effective temperature}},\ }\href {https://doi.org/10.1103/PhysRevE.90.012111} {\bibfield  {journal} {\bibinfo  {journal} {Phys. Rev. E}\ }\textbf {\bibinfo {volume} {90}},\ \bibinfo {pages} {012111} (\bibinfo {year} {2014})}\BibitemShut {NoStop}%
\bibitem [{\citenamefont {Martin}\ \emph {et~al.}(2021)\citenamefont {Martin}, \citenamefont {O'Byrne}, \citenamefont {Cates}, \citenamefont {Fodor}, \citenamefont {Nardini}, \citenamefont {Tailleur},\ and\ \citenamefont {van Wijland}}]{MartinPRE2021}%
  \BibitemOpen
  \bibfield  {author} {\bibinfo {author} {\bibfnamefont {D.}~\bibnamefont {Martin}}, \bibinfo {author} {\bibfnamefont {J.}~\bibnamefont {O'Byrne}}, \bibinfo {author} {\bibfnamefont {M.~E.}\ \bibnamefont {Cates}}, \bibinfo {author} {\bibfnamefont {{\'{E}}.}~\bibnamefont {Fodor}}, \bibinfo {author} {\bibfnamefont {C.}~\bibnamefont {Nardini}}, \bibinfo {author} {\bibfnamefont {J.}~\bibnamefont {Tailleur}},\ and\ \bibinfo {author} {\bibfnamefont {F.}~\bibnamefont {van Wijland}},\ }\bibfield  {title} {\bibinfo {title} {{Statistical mechanics of active Ornstein-Uhlenbeck particles}},\ }\href {https://doi.org/10.1103/PhysRevE.103.032607} {\bibfield  {journal} {\bibinfo  {journal} {Phys. Rev. E}\ }\textbf {\bibinfo {volume} {103}},\ \bibinfo {pages} {032607} (\bibinfo {year} {2021})}\BibitemShut {NoStop}%
\bibitem [{\citenamefont {Solon}\ and\ \citenamefont {Tailleur}(2013)}]{SolonPRL2013}%
  \BibitemOpen
  \bibfield  {author} {\bibinfo {author} {\bibfnamefont {A.~P.}\ \bibnamefont {Solon}}\ and\ \bibinfo {author} {\bibfnamefont {J.}~\bibnamefont {Tailleur}},\ }\bibfield  {title} {\bibinfo {title} {{Revisiting the Flocking Transition Using Active Spins}},\ }\href {https://doi.org/10.1103/PhysRevLett.111.078101} {\bibfield  {journal} {\bibinfo  {journal} {Phys. Rev. Lett.}\ }\textbf {\bibinfo {volume} {111}},\ \bibinfo {pages} {078101} (\bibinfo {year} {2013})}\BibitemShut {NoStop}%
\bibitem [{\citenamefont {Solon}\ and\ \citenamefont {Tailleur}(2015)}]{SolonPRE2015}%
  \BibitemOpen
  \bibfield  {author} {\bibinfo {author} {\bibfnamefont {A.~P.}\ \bibnamefont {Solon}}\ and\ \bibinfo {author} {\bibfnamefont {J.}~\bibnamefont {Tailleur}},\ }\bibfield  {title} {\bibinfo {title} {{Flocking with discrete symmetry: The two-dimensional active Ising model}},\ }\href {https://doi.org/10.1103/PhysRevE.92.042119} {\bibfield  {journal} {\bibinfo  {journal} {Phys. Rev. E}\ }\textbf {\bibinfo {volume} {92}},\ \bibinfo {pages} {042119} (\bibinfo {year} {2015})}\BibitemShut {NoStop}%
\bibitem [{\citenamefont {Benvegnen}\ \emph {et~al.}(2023)\citenamefont {Benvegnen}, \citenamefont {Granek}, \citenamefont {Ro}, \citenamefont {Yaacoby}, \citenamefont {Chat\'e}, \citenamefont {Kafri}, \citenamefont {Mukamel}, \citenamefont {Solon},\ and\ \citenamefont {Tailleur}}]{BenvegnenPRL2023}%
  \BibitemOpen
  \bibfield  {author} {\bibinfo {author} {\bibfnamefont {B.}~\bibnamefont {Benvegnen}}, \bibinfo {author} {\bibfnamefont {O.}~\bibnamefont {Granek}}, \bibinfo {author} {\bibfnamefont {S.}~\bibnamefont {Ro}}, \bibinfo {author} {\bibfnamefont {R.}~\bibnamefont {Yaacoby}}, \bibinfo {author} {\bibfnamefont {H.}~\bibnamefont {Chat\'e}}, \bibinfo {author} {\bibfnamefont {Y.}~\bibnamefont {Kafri}}, \bibinfo {author} {\bibfnamefont {D.}~\bibnamefont {Mukamel}}, \bibinfo {author} {\bibfnamefont {A.}~\bibnamefont {Solon}},\ and\ \bibinfo {author} {\bibfnamefont {J.}~\bibnamefont {Tailleur}},\ }\bibfield  {title} {\bibinfo {title} {{Metastability of Discrete-Symmetry Flocks}},\ }\href {https://doi.org/10.1103/PhysRevLett.131.218301} {\bibfield  {journal} {\bibinfo  {journal} {Phys. Rev. Lett.}\ }\textbf {\bibinfo {volume} {131}},\ \bibinfo {pages} {218301} (\bibinfo {year} {2023})}\BibitemShut {NoStop}%
\bibitem [{\citenamefont {Woo}\ and\ \citenamefont {Noh}(2024)}]{WooPRL2024}%
  \BibitemOpen
  \bibfield  {author} {\bibinfo {author} {\bibfnamefont {C.-U.}\ \bibnamefont {Woo}}\ and\ \bibinfo {author} {\bibfnamefont {J.~D.}\ \bibnamefont {Noh}},\ }\bibfield  {title} {\bibinfo {title} {{Motility-Induced Pinning in Flocking System with Discrete Symmetry}},\ }\href {https://doi.org/10.1103/PhysRevLett.133.188301} {\bibfield  {journal} {\bibinfo  {journal} {Phys. Rev. Lett.}\ }\textbf {\bibinfo {volume} {133}},\ \bibinfo {pages} {188301} (\bibinfo {year} {2024})}\BibitemShut {NoStop}%
\bibitem [{\citenamefont {Chatterjee}\ \emph {et~al.}(2020)\citenamefont {Chatterjee}, \citenamefont {Mangeat}, \citenamefont {Paul},\ and\ \citenamefont {Rieger}}]{ChatterjeeEPL2020}%
  \BibitemOpen
  \bibfield  {author} {\bibinfo {author} {\bibfnamefont {S.}~\bibnamefont {Chatterjee}}, \bibinfo {author} {\bibfnamefont {M.}~\bibnamefont {Mangeat}}, \bibinfo {author} {\bibfnamefont {R.}~\bibnamefont {Paul}},\ and\ \bibinfo {author} {\bibfnamefont {H.}~\bibnamefont {Rieger}},\ }\bibfield  {title} {\bibinfo {title} {{Flocking and reorientation transition in the 4-state active Potts model}},\ }\href {https://doi.org/10.1209/0295-5075/130/66001} {\bibfield  {journal} {\bibinfo  {journal} {Europhys. Lett.}\ }\textbf {\bibinfo {volume} {130}},\ \bibinfo {pages} {66001} (\bibinfo {year} {2020})}\BibitemShut {NoStop}%
\bibitem [{\citenamefont {Mangeat}\ \emph {et~al.}(2020)\citenamefont {Mangeat}, \citenamefont {Chatterjee}, \citenamefont {Paul},\ and\ \citenamefont {Rieger}}]{MangeatPRE2020}%
  \BibitemOpen
  \bibfield  {author} {\bibinfo {author} {\bibfnamefont {M.}~\bibnamefont {Mangeat}}, \bibinfo {author} {\bibfnamefont {S.}~\bibnamefont {Chatterjee}}, \bibinfo {author} {\bibfnamefont {R.}~\bibnamefont {Paul}},\ and\ \bibinfo {author} {\bibfnamefont {H.}~\bibnamefont {Rieger}},\ }\bibfield  {title} {\bibinfo {title} {{Flocking with a $q$-fold discrete symmetry: Band-to-lane transition in the active Potts model}},\ }\href {https://doi.org/10.1103/PhysRevE.102.042601} {\bibfield  {journal} {\bibinfo  {journal} {Phys. Rev. E}\ }\textbf {\bibinfo {volume} {102}},\ \bibinfo {pages} {042601} (\bibinfo {year} {2020})}\BibitemShut {NoStop}%
\bibitem [{\citenamefont {Solon}\ \emph {et~al.}(2022)\citenamefont {Solon}, \citenamefont {Chat\'e}, \citenamefont {Toner},\ and\ \citenamefont {Tailleur}}]{SolonPRL2022}%
  \BibitemOpen
  \bibfield  {author} {\bibinfo {author} {\bibfnamefont {A.}~\bibnamefont {Solon}}, \bibinfo {author} {\bibfnamefont {H.}~\bibnamefont {Chat\'e}}, \bibinfo {author} {\bibfnamefont {J.}~\bibnamefont {Toner}},\ and\ \bibinfo {author} {\bibfnamefont {J.}~\bibnamefont {Tailleur}},\ }\bibfield  {title} {\bibinfo {title} {{Susceptibility of Polar Flocks to Spatial Anisotropy}},\ }\href {https://doi.org/10.1103/PhysRevLett.128.208004} {\bibfield  {journal} {\bibinfo  {journal} {Phys. Rev. Lett.}\ }\textbf {\bibinfo {volume} {128}},\ \bibinfo {pages} {208004} (\bibinfo {year} {2022})}\BibitemShut {NoStop}%
\bibitem [{\citenamefont {Chatterjee}\ \emph {et~al.}(2022)\citenamefont {Chatterjee}, \citenamefont {Mangeat},\ and\ \citenamefont {Rieger}}]{ChatterjeeEPL2022}%
  \BibitemOpen
  \bibfield  {author} {\bibinfo {author} {\bibfnamefont {S.}~\bibnamefont {Chatterjee}}, \bibinfo {author} {\bibfnamefont {M.}~\bibnamefont {Mangeat}},\ and\ \bibinfo {author} {\bibfnamefont {H.}~\bibnamefont {Rieger}},\ }\bibfield  {title} {\bibinfo {title} {{Polar flocks with discretized directions: The active clock model approaching the Vicsek model}},\ }\href {https://doi.org/10.1209/0295-5075/ac6e4b} {\bibfield  {journal} {\bibinfo  {journal} {Europhys. Lett.}\ }\textbf {\bibinfo {volume} {138}},\ \bibinfo {pages} {41001} (\bibinfo {year} {2022})}\BibitemShut {NoStop}%
\bibitem [{\citenamefont {Peruani}\ \emph {et~al.}(2011)\citenamefont {Peruani}, \citenamefont {Klauss}, \citenamefont {Deutsch},\ and\ \citenamefont {Voss-Boehme}}]{PeruaniPRL2011}%
  \BibitemOpen
  \bibfield  {author} {\bibinfo {author} {\bibfnamefont {F.}~\bibnamefont {Peruani}}, \bibinfo {author} {\bibfnamefont {T.}~\bibnamefont {Klauss}}, \bibinfo {author} {\bibfnamefont {A.}~\bibnamefont {Deutsch}},\ and\ \bibinfo {author} {\bibfnamefont {A.}~\bibnamefont {Voss-Boehme}},\ }\bibfield  {title} {\bibinfo {title} {{Traffic Jams, Gliders, and Bands in the Quest for Collective Motion of Self-Propelled Particles}},\ }\href {https://doi.org/10.1103/PhysRevLett.106.128101} {\bibfield  {journal} {\bibinfo  {journal} {Phys. Rev. Lett.}\ }\textbf {\bibinfo {volume} {106}},\ \bibinfo {pages} {128101} (\bibinfo {year} {2011})}\BibitemShut {NoStop}%
\bibitem [{\citenamefont {Rosembach}\ \emph {et~al.}(2024)\citenamefont {Rosembach}, \citenamefont {Dias},\ and\ \citenamefont {Dickman}}]{RosembachPRE2024}%
  \BibitemOpen
  \bibfield  {author} {\bibinfo {author} {\bibfnamefont {T.~V.}\ \bibnamefont {Rosembach}}, \bibinfo {author} {\bibfnamefont {A.~L.~N.}\ \bibnamefont {Dias}},\ and\ \bibinfo {author} {\bibfnamefont {R.}~\bibnamefont {Dickman}},\ }\bibfield  {title} {\bibinfo {title} {{Three-state active lattice gas: A discrete Vicsek-like model with excluded volume}},\ }\href {https://doi.org/10.1103/PhysRevE.110.014109} {\bibfield  {journal} {\bibinfo  {journal} {Phys. Rev. E}\ }\textbf {\bibinfo {volume} {110}},\ \bibinfo {pages} {014109} (\bibinfo {year} {2024})}\BibitemShut {NoStop}%
\bibitem [{\citenamefont {Thompson}\ \emph {et~al.}(2011)\citenamefont {Thompson}, \citenamefont {Tailleur}, \citenamefont {Cates},\ and\ \citenamefont {Blythe}}]{ThompsonJSTAT2011}%
  \BibitemOpen
  \bibfield  {author} {\bibinfo {author} {\bibfnamefont {A.~G.}\ \bibnamefont {Thompson}}, \bibinfo {author} {\bibfnamefont {J.}~\bibnamefont {Tailleur}}, \bibinfo {author} {\bibfnamefont {M.~E.}\ \bibnamefont {Cates}},\ and\ \bibinfo {author} {\bibfnamefont {R.~A.}\ \bibnamefont {Blythe}},\ }\bibfield  {title} {\bibinfo {title} {{Lattice models of nonequilibrium bacterial dynamics}},\ }\href {https://doi.org/10.1088/1742-5468/2011/02/P02029} {\bibfield  {journal} {\bibinfo  {journal} {J. Stat. Mech.: Theor. Exp.}\ }\textbf {\bibinfo {volume} {2011}},\ \bibinfo {pages} {P02029} (\bibinfo {year} {2011})}\BibitemShut {NoStop}%
\bibitem [{\citenamefont {Whitelam}\ \emph {et~al.}(2018)\citenamefont {Whitelam}, \citenamefont {Klymko},\ and\ \citenamefont {Mandal}}]{WhitelamJCP2018}%
  \BibitemOpen
  \bibfield  {author} {\bibinfo {author} {\bibfnamefont {S.}~\bibnamefont {Whitelam}}, \bibinfo {author} {\bibfnamefont {K.}~\bibnamefont {Klymko}},\ and\ \bibinfo {author} {\bibfnamefont {D.}~\bibnamefont {Mandal}},\ }\bibfield  {title} {\bibinfo {title} {{Phase separation and large deviations of lattice active matter}},\ }\href {https://doi.org/10.1063/1.5023403} {\bibfield  {journal} {\bibinfo  {journal} {J. Chem. Phys.}\ }\textbf {\bibinfo {volume} {148}},\ \bibinfo {pages} {154902} (\bibinfo {year} {2018})}\BibitemShut {NoStop}%
\bibitem [{\citenamefont {Partridge}\ and\ \citenamefont {Lee}(2019)}]{PartridgePRL2019}%
  \BibitemOpen
  \bibfield  {author} {\bibinfo {author} {\bibfnamefont {B.}~\bibnamefont {Partridge}}\ and\ \bibinfo {author} {\bibfnamefont {C.~F.}\ \bibnamefont {Lee}},\ }\bibfield  {title} {\bibinfo {title} {{Critical Motility-Induced Phase Separation Belongs to the Ising Universality Class}},\ }\href {https://doi.org/10.1103/PhysRevLett.123.068002} {\bibfield  {journal} {\bibinfo  {journal} {Phys. Rev. Lett.}\ }\textbf {\bibinfo {volume} {123}},\ \bibinfo {pages} {068002} (\bibinfo {year} {2019})}\BibitemShut {NoStop}%
\bibitem [{\citenamefont {Shi}\ \emph {et~al.}(2020)\citenamefont {Shi}, \citenamefont {Fausti}, \citenamefont {Chat\'e}, \citenamefont {Nardini},\ and\ \citenamefont {Solon}}]{ShiPRL2020}%
  \BibitemOpen
  \bibfield  {author} {\bibinfo {author} {\bibfnamefont {X.-q.}\ \bibnamefont {Shi}}, \bibinfo {author} {\bibfnamefont {G.}~\bibnamefont {Fausti}}, \bibinfo {author} {\bibfnamefont {H.}~\bibnamefont {Chat\'e}}, \bibinfo {author} {\bibfnamefont {C.}~\bibnamefont {Nardini}},\ and\ \bibinfo {author} {\bibfnamefont {A.}~\bibnamefont {Solon}},\ }\bibfield  {title} {\bibinfo {title} {{Self-Organized Critical Coexistence Phase in Repulsive Active Particles}},\ }\href {https://doi.org/10.1103/PhysRevLett.125.168001} {\bibfield  {journal} {\bibinfo  {journal} {Phys. Rev. Lett.}\ }\textbf {\bibinfo {volume} {125}},\ \bibinfo {pages} {168001} (\bibinfo {year} {2020})}\BibitemShut {NoStop}%
\bibitem [{\citenamefont {Sep\'ulveda}\ and\ \citenamefont {Soto}(2017)}]{SepulvedaPRL2017}%
  \BibitemOpen
  \bibfield  {author} {\bibinfo {author} {\bibfnamefont {N.}~\bibnamefont {Sep\'ulveda}}\ and\ \bibinfo {author} {\bibfnamefont {R.}~\bibnamefont {Soto}},\ }\bibfield  {title} {\bibinfo {title} {{Wetting Transitions Displayed by Persistent Active Particles}},\ }\href {https://doi.org/10.1103/PhysRevLett.119.078001} {\bibfield  {journal} {\bibinfo  {journal} {Phys. Rev. Lett.}\ }\textbf {\bibinfo {volume} {119}},\ \bibinfo {pages} {078001} (\bibinfo {year} {2017})}\BibitemShut {NoStop}%
\bibitem [{\citenamefont {Agranov}\ \emph {et~al.}(2022)\citenamefont {Agranov}, \citenamefont {Cates},\ and\ \citenamefont {Jack}}]{AgranovJSTAT2022}%
  \BibitemOpen
  \bibfield  {author} {\bibinfo {author} {\bibfnamefont {T.}~\bibnamefont {Agranov}}, \bibinfo {author} {\bibfnamefont {M.~E.}\ \bibnamefont {Cates}},\ and\ \bibinfo {author} {\bibfnamefont {R.~L.}\ \bibnamefont {Jack}},\ }\bibfield  {title} {\bibinfo {title} {{Entropy production and its large deviations in an active lattice gas}},\ }\href {https://doi.org/10.1088/1742-5468/aca0eb} {\bibfield  {journal} {\bibinfo  {journal} {J. Stat. Mech.: Theor. Exp.}\ }\textbf {\bibinfo {volume} {2022}},\ \bibinfo {pages} {123201} (\bibinfo {year} {2022})}\BibitemShut {NoStop}%
\bibitem [{\citenamefont {Agranov}\ \emph {et~al.}(2024)\citenamefont {Agranov}, \citenamefont {Jack}, \citenamefont {Cates},\ and\ \citenamefont {Fodor}}]{AgranovNJP2024}%
  \BibitemOpen
  \bibfield  {author} {\bibinfo {author} {\bibfnamefont {T.}~\bibnamefont {Agranov}}, \bibinfo {author} {\bibfnamefont {R.~L.}\ \bibnamefont {Jack}}, \bibinfo {author} {\bibfnamefont {M.~E.}\ \bibnamefont {Cates}},\ and\ \bibinfo {author} {\bibfnamefont {{\'{E}}.}~\bibnamefont {Fodor}},\ }\bibfield  {title} {\bibinfo {title} {{Thermodynamically consistent flocking: from discontinuous to continuous transition}},\ }\href {https://doi.org/10.1088/1367-2630/ad4dd6} {\bibfield  {journal} {\bibinfo  {journal} {New J. Phys.}\ }\textbf {\bibinfo {volume} {26}},\ \bibinfo {pages} {063006} (\bibinfo {year} {2024})}\BibitemShut {NoStop}%
\bibitem [{\citenamefont {Katz}\ \emph {et~al.}(1984)\citenamefont {Katz}, \citenamefont {Lebowitz},\ and\ \citenamefont {Spohn}}]{KatzJSP1984}%
  \BibitemOpen
  \bibfield  {author} {\bibinfo {author} {\bibfnamefont {S.}~\bibnamefont {Katz}}, \bibinfo {author} {\bibfnamefont {J.~L.}\ \bibnamefont {Lebowitz}},\ and\ \bibinfo {author} {\bibfnamefont {H.}~\bibnamefont {Spohn}},\ }\bibfield  {title} {\bibinfo {title} {{Nonequilibrium steady states of stochastic lattice gas models of fast ionic conductors}},\ }\href {https://doi.org/10.1007/bf01018556} {\bibfield  {journal} {\bibinfo  {journal} {J. Stat. Phys.}\ }\textbf {\bibinfo {volume} {34}},\ \bibinfo {pages} {497 } (\bibinfo {year} {1984})}\BibitemShut {NoStop}%
\bibitem [{\citenamefont {Maes}(2021)}]{MaesSciPost2021}%
  \BibitemOpen
  \bibfield  {author} {\bibinfo {author} {\bibfnamefont {C.}~\bibnamefont {Maes}},\ }\bibfield  {title} {\bibinfo {title} {{Local detailed balance}},\ }\href {https://doi.org/10.21468/SciPostPhysLectNotes.32} {\bibfield  {journal} {\bibinfo  {journal} {SciPost Phys. Lect. Notes}\ ,\ \bibinfo {pages} {32}} (\bibinfo {year} {2021})}\BibitemShut {NoStop}%
\bibitem [{\citenamefont {Seifert}(2005)}]{SeifertPRL2005}%
  \BibitemOpen
  \bibfield  {author} {\bibinfo {author} {\bibfnamefont {U.}~\bibnamefont {Seifert}},\ }\bibfield  {title} {\bibinfo {title} {{Entropy Production along a Stochastic Trajectory and an Integral Fluctuation Theorem}},\ }\href {http://link.aps.org/doi/10.1103/PhysRevLett.95.040602} {\bibfield  {journal} {\bibinfo  {journal} {Phys. Rev. Lett.}\ }\textbf {\bibinfo {volume} {95}},\ \bibinfo {pages} {040602} (\bibinfo {year} {2005})}\BibitemShut {NoStop}%
\bibitem [{\citenamefont {Kawasaki}(1972)}]{Kawasaki1972}%
  \BibitemOpen
  \bibfield  {author} {\bibinfo {author} {\bibfnamefont {K.}~\bibnamefont {Kawasaki}},\ }\bibfield  {title} {\bibinfo {title} {{Kinetics of Ising Models}},\ }in\ \href@noop {} {\emph {\bibinfo {booktitle} {{Phase Transitions and Critical Phenomena}}}},\ Vol.~\bibinfo {volume} {2},\ \bibinfo {editor} {edited by\ \bibinfo {editor} {\bibfnamefont {C.}~\bibnamefont {Domb}}\ and\ \bibinfo {editor} {\bibfnamefont {M.}~\bibnamefont {Green}}}\ (\bibinfo  {publisher} {Academic},\ \bibinfo {address} {London},\ \bibinfo {year} {1972})\BibitemShut {NoStop}%
\bibitem [{\citenamefont {Glauber}(1963)}]{GlauberJMP1963}%
  \BibitemOpen
  \bibfield  {author} {\bibinfo {author} {\bibfnamefont {R.~J.}\ \bibnamefont {Glauber}},\ }\bibfield  {title} {\bibinfo {title} {{Time‐Dependent Statistics of the Ising Model}},\ }\href {https://doi.org/10.1063/1.1703954} {\bibfield  {journal} {\bibinfo  {journal} {J. Math. Phys.}\ }\textbf {\bibinfo {volume} {4}},\ \bibinfo {pages} {294} (\bibinfo {year} {1963})}\BibitemShut {NoStop}%
\bibitem [{\citenamefont {Yao}\ and\ \citenamefont {Jack}(2024)}]{Yao2024}%
  \BibitemOpen
  \bibfield  {author} {\bibinfo {author} {\bibfnamefont {L.}~\bibnamefont {Yao}}\ and\ \bibinfo {author} {\bibfnamefont {R.~L.}\ \bibnamefont {Jack}},\ }\href {https://arxiv.org/abs/2412.04450} {} (\bibinfo {year} {2024}),\ \Eprint {https://arxiv.org/abs/2412.04450} {arXiv:2412.04450 [cond-mat.soft]} \BibitemShut {NoStop}%
\bibitem [{\citenamefont {Kloeden}\ and\ \citenamefont {Platen}(1999)}]{KloedenBook1999}%
  \BibitemOpen
  \bibfield  {author} {\bibinfo {author} {\bibfnamefont {P.~E.}\ \bibnamefont {Kloeden}}\ and\ \bibinfo {author} {\bibfnamefont {E.}~\bibnamefont {Platen}},\ }\href@noop {} {\emph {\bibinfo {title} {{Numerical Solution of Stochastic Differential Equations}}}}\ (\bibinfo  {publisher} {Springer},\ \bibinfo {year} {1999})\BibitemShut {NoStop}%
\bibitem [{\citenamefont {Asmussen}\ and\ \citenamefont {Glynn}(2007)}]{AsmussenBook2007}%
  \BibitemOpen
  \bibfield  {author} {\bibinfo {author} {\bibfnamefont {S.}~\bibnamefont {Asmussen}}\ and\ \bibinfo {author} {\bibfnamefont {P.~W.}\ \bibnamefont {Glynn}},\ }\href@noop {} {\emph {\bibinfo {title} {{Stochastic Simulation: Algorithms and Analysis}}}}\ (\bibinfo  {publisher} {Springer},\ \bibinfo {year} {2007})\BibitemShut {NoStop}%
\bibitem [{\citenamefont {Gardiner}(2009)}]{GardinerBook2009}%
  \BibitemOpen
  \bibfield  {author} {\bibinfo {author} {\bibfnamefont {C.}~\bibnamefont {Gardiner}},\ }\href@noop {} {\emph {\bibinfo {title} {{Stochastic Methods: A Handbook for the Natural and Social Sciences}}}}\ (\bibinfo  {publisher} {Springer},\ \bibinfo {year} {2009})\BibitemShut {NoStop}%
\bibitem [{\citenamefont {Solon}\ \emph {et~al.}(2015)\citenamefont {Solon}, \citenamefont {Fily}, \citenamefont {Baskaran}, \citenamefont {Cates}, \citenamefont {Kafri}, \citenamefont {Kardar},\ and\ \citenamefont {Tailleur}}]{SolonNP2015}%
  \BibitemOpen
  \bibfield  {author} {\bibinfo {author} {\bibfnamefont {A.~P.}\ \bibnamefont {Solon}}, \bibinfo {author} {\bibfnamefont {Y.}~\bibnamefont {Fily}}, \bibinfo {author} {\bibfnamefont {A.}~\bibnamefont {Baskaran}}, \bibinfo {author} {\bibfnamefont {M.~E.}\ \bibnamefont {Cates}}, \bibinfo {author} {\bibfnamefont {Y.}~\bibnamefont {Kafri}}, \bibinfo {author} {\bibfnamefont {M.}~\bibnamefont {Kardar}},\ and\ \bibinfo {author} {\bibfnamefont {J.}~\bibnamefont {Tailleur}},\ }\bibfield  {title} {\bibinfo {title} {{Pressure is not a state function for generic active fluids}},\ }\href {https://doi.org/10.1038/nphys3377} {\bibfield  {journal} {\bibinfo  {journal} {Nat. Phys.}\ }\textbf {\bibinfo {volume} {11}},\ \bibinfo {pages} {673} (\bibinfo {year} {2015})}\BibitemShut {NoStop}%
\bibitem [{cod()}]{code}%
  \BibitemOpen
  \href@noop {} {}\bibinfo {note} {Available online at \url{https://github.com/Ki-Won/Active-particle-lattice-sim}.}\BibitemShut {Stop}%
\end{thebibliography}%

\end{document}